\documentclass[aps,prd,superscriptaddress,nofootinbib,11pt]{revtex4}
\usepackage[english]{babel}
\usepackage[utf8]{inputenc}
\usepackage{graphicx}   
\usepackage{slashed}
\usepackage{epstopdf}
\usepackage{verbatim}   
\usepackage{color}      
\usepackage{subfigure}  
\usepackage{multirow}
\usepackage{hyperref}   
\usepackage{float}
\usepackage{epsfig,rotating}
\usepackage{amsmath,amssymb}
\usepackage{dsfont}
\usepackage{slashed}
\restylefloat{table}
\numberwithin{equation}{section}

\newcommand{\vx}{\vec{x}}
\newcommand{\vp}{\vec{p}}
\newcommand{\vq}{\vec{q}}
\newcommand{\vk}{\vec{k}}

\newcommand{\vy}{\vec{y}}

\newcommand{\be}{\begin{equation}}
\newcommand{\ee}{\end{equation}}
\newcommand{\bea}{\begin{eqnarray}}
\newcommand{\eea}{\end{eqnarray}}

\newcommand{\ket}[1]{|#1\rangle}
\newcommand{\bra}[1]{\langle#1|}

\begin{document}
\title{ Dynamical axion quasiparticles: an open quantum system.}

\author{Daniel Boyanovsky}
\email{boyan@pitt.edu} \affiliation{Department of Physics and
Astronomy, University of Pittsburgh, Pittsburgh, PA 15260}

 \date{\today}

\begin{abstract}
We study the non-equilibrium dynamics of emergent dynamical axion quasiparticles (DAQ) coupled to a photon bath in equilibrium via a Chern-Simons term as a quantum open system. A quantum master equation (QME) is derived up to  second order in this coupling implementing only a \emph{partial} Markov approximation, allowing time dependent rates in the Lindblad (QME). These are determined by the equilibrium correlation functions of the Chern-Simons density, and their time dependence allows us to explore transient dynamics in coherences and population: the formation of the quasiparticle on short time scales and its decay, and the build-up of population with an effective time dependent rate. Early time evolution features quantum \emph{anti} Zeno dynamics with enhanced quasiparticle decay  and population growth. These phenomena describe transient violations of Fermi's Golden rule and of \emph{detailed balance}, and are distinct \emph{non-Markovian} effects   directly related to the spectral density of the Chern-Simons correlators. We obtain the equation of motion of coherent (DAQ) condensates both with the (QME) and with quantum many body linear response establishing a direct bridge between both methods. As a corollary we obtain the expectation value of the Chern-Simons density \emph{induced} by a (DAQ) condensate in linear response, the topological susceptibility is shown to be proportional to the (DAQ) many body self-energy. We provide a Feynman diagram-based interpretation of approximations invoked in the (QME) and   corrections from system-bath correlations in higher order.

\end{abstract}

\maketitle

\section{Introduction}
 Dynamical axion quasiparticles (DAQ) emerge as collective excitations in magnetic topological insulators that break parity and time reversal invariance\cite{xiao,rundong,jing,nomura,narang}, charge density waves in Weyl semimetals with broken parity and time reversal invariance\cite{gooth,gos,yu,mottola}, or in multilayered metamaterials\cite{wilczekshapo}, and phononic induced axions in Dirac insulators\cite{nabil}. They are the condensed matter analog of cosmological axions,   originally proposed as a solution of the strong CP problem in Quantum Chromodynamics\cite{PQ,weinaxion,wil} and is a compelling dark matter candidate\cite{marsh,sikivie1,press,abbott,fischler}. An important aspect that characterizes the (DAQ) field  in common with  the cosmic axion is  its   coupling to electromagnetism   via a topological Chern-Simons term $ \propto \epsilon_{\mu\nu\rho\sigma}F^{\mu\nu} F^{\rho \sigma} \propto \vec{E}\cdot\vec{B}$,  giving rise to novel phenomena:    axion electrodynamics\cite{wilczekaxion}. This particular coupling leads to topological magnetoelectric effects such as Faraday and Kerr rotations\cite{liang,tse,ahn},  the hybridization of photons and axions, namely \emph{axionic polaritons}\cite{rundong,zhu,shio} in presence of an external magnetic field, and the possibility of  hybridization of (DAQ) with the cosmological axion\cite{boyhybrid}. Axion-photon mixing is also at the heart of the Primakoff effect that is proposed as an experimental method of detection of the dark matter axion\cite{marsh2,chig}.  Dynamical axion quasiparticles are manifest as coherent oscillations of the (DAQ) field that break parity and time reversal symmetry.  Recently the observation of such coherent oscillation   induced by an antiferromagnetic magnon   has been reported in two dimensional ($MnBi_2Te_4$)\cite{jianq} and interpreted as plausible and consistent    evidence for a dynamical axion quasiparticle. This observation  bolsters the case for a deeper understanding of axion quasiparticles in three dimensions, where the vanishing  of the bulk gap at the surface of magnetic topological insulators yield an enhancement of the fluctuations of the dynamical axion field near the surface providing a possible platform for their detection\cite{zaletel}. The coupling of (DAQ) to electromagnetism via the Chern-simons term leads to the decay of the emergent quasiparticle into two photons providing a possible observational avenue via multiphoton correlations\cite{olivia}. Parity and time reversal breaking is imprinted as a telltale in the two-photon final state from (DAQ) decay in the form of Bell-type correlated pairs \emph{hyperentangled} in momentum and polarization amenable of being probed via two-photon interferometry\cite{boyphot}.

 \vspace{1mm}

 \textbf{Motivation, objectives and brief summary of results:}
 The observation of damped coherent oscillations induced by antiferromagnetic magnons in two dimensional ($MnBi_2Te_4$), interpreted as evidence of   dynamical axion quasiparticles\cite{jianq}, motivates our study of  the non-equilibrium dynamics of emergent (DAQ) coupled via the Chern-Simons term to a photon bath in equilibrium as a \emph{quantum open system }\cite{breuer,zoeller,carmichael} in three dimensions. The coupling of (DAQ) to electromagnetism via a topological Chern-Simons term is the hallmark of axion electrodynamics\cite{wilczekaxion} and a distinguishing feature of these collective excitations whose existence implies the breaking of parity and time reversal invariance.

 Our objectives in this study are to explore non-equilibrium aspects of (DAQ)   as a consequence of their coupling to a photon bath via a Chern-Simons term, in particular the non-equilibrium evolution of a coherent (DAQ) condensate, which is the focus of the experimental study of reference\cite{jianq}, the dynamics of formation of the quasiparticle,   decoherence and population kinetics. Our study focuses  on transient dynamics that may be experimentally accessible with current methods of ultrafast spectroscopy and that could be potential experimental probes of the  (DAQ) coupling to photons via the Chern-Simons term.

 With this aim, we derive a quantum master equation (QME) by evolving an initial density matrix and tracing out the (photon) bath degrees of freedom up to second order in the (DAQ)-photon coupling. Invoking ubiquitous approximations\cite{breuer,zoeller,carmichael} we obtain the quantum master equation for the (DAQ) reduced density matrix in Lindblad form\cite{lin,gori,pearle,linme}. However we do \emph{not} invoke a full Markov approximation, keeping the   time dependence of the coefficient functions of the Lindbladians which are determined by the equilibrium correlations of the Chern-Simons density, rather than taking their infinite time limit. This allows us to explore non-Markovian (memory)\cite{nonmarkov,vega} processes to understand   transient phenomena. While the derivation of the (QME)   for the case of (DAQ) photon coupling up to second order in the coupling is similar to that obtained in ref.\cite{cosmoqme}  in a different context, it is extended and complemented by  a detailed comparison to  many body linear response,  thereby establishing a bridge between the (QME) and quantum many body theory. This correspondence  allows us   to confirm the validity of the main approximations \emph{up to second order in the coupling}, and to provide a Feynman diagram-based interpretation of factorization and emergence of system-bath correlations in higher orders.   Therefore, by establishing the bridge between the (QME) and quantum many body theory, we provide a sound justification of the validity of the various approximations up to second order in the coupling.    The general   validity of the approximations leading to  the Lindblad form of the (QME)  has been   critically scrutinized  recently in ref.\cite{linme}.

  \vspace{1mm}

  \textbf{Summary of results:}

  \vspace{1mm}
  From the (QME) we obtain the time evolution of the (DAQ) condensate, coherences and population (occupation number). In all cases, we find important transient dynamics  associated with the formation of the quasiparticle,  \textbf{i:)} early time evolution that is the precursor  of  the  quantum Zeno effect\cite{fonda,zeno1,zeno2} but is manifest as the quantum \emph{anti}-Zeno effect with enhanced decay\cite{zeno3} and population build-up, \textbf{ii:)} violations of Fermi's golden rule and of detailed balance. These transient effects are  missed altogether  by the Markov approximation yet   may be amenable to  be explored  via ultrafast spectroscopy. The quantum  Zeno effect is a telltale of non-Markovian dynamics that has already been confirmed experimentally with cold ${}^9Be^+$ atoms confined in  a Penning trap\cite{itano}, and ultrafast interferometry has been implemented to study the real time dynamics of a quasiparticle in the case of impurities coupled to a Fermi sea\cite{cetina,viva}. Characteristic properties of the spectral density of Chern-Simons correlators lead to transient early time anti-Zeno\cite{zeno3} dynamics: enhanced decay of the quasiparticle, rather than inhibited decay, and faster population build-up as compared to Markovian dynamics.

  We recognize a hierarchy of time scales,  the shortest determined by an upper cutoff of the bath spectral density is identified with the formation time of the quasiparticle. A wide separation of scales allows us to obtain the effective non-equilibrium dynamics of the quasiparticle, coherences and population  on longer time scales, including  early time transients that feature  quantum anti-Zeno dynamics  which imply a short time violation of Fermi's Golden rule and, crucially,  of detailed balance.

 We obtain the  equation of motion of a coherent (DAQ) condensate from the \emph{time non-local} (QME) obtained \emph{without invoking the Markov and rotating wave approximation}, the non-locality in the equation of motion is recognized in terms of the many body retarded (DAQ) self-energy.  We  complement the (QME) approach  with an alternative derivation of the equation of motion of the coherent (DAQ) condensate implementing quantum many body linear response and find complete agreement with that obtained from the \emph{time non-local } (QME). This alternative derivation illuminates the physical content of some of the approximations in the (QME) which along with a multitime analysis of the solutions  confirms the transient dynamics and provide a benchmark to test the reliability of the various approximations. This complementary   approach provides a direct bridge between the (QME) and  quantum many body physics. As a corollary of the many body  linear response approach we obtain the expectation value of the Chern-Simons density \emph{induced} by a coherent (DAQ) condensate, and relate the topological susceptibility to the   (DAQ) retarded self-energy. This correspondence with quantum many body physics allows us to provide a Feynman diagram-based understanding of important approximations leading to the reduced density matrix and an interpretation of the build-up of system-bath correlations in higher orders.

 The article is organized as follows: in section (\ref{sec:daq}) we summarize the effective field theory of (DAQ)\cite{xiao,rundong,jing,nomura,narang,mottola} coupled to the electromagnetic field via a Chern-Simons term, derive the (QME) up to second order in the Chern-Simons coupling, discussing the various approximations invoked in the literature and obtain the Lindblad form, but with coefficient functions that depend explicitly on time and are determined by correlations of the Chern-Simons density. In this section we obtain the equations of motion for population and coherences. In section (\ref{sec:transient}) we study transient non-Markovian dynamics as a consequence of allowing the coefficients of the Lindblad dissipator to depend on time. We recognize a hierarchy of time scales determined by  the spectral properties of the Chern-Simons correlation functions. The shortest is identified with the time scale of formation of the quasiparticle yielding the ``quasiparticle residue''. The evolution of the quasiparticle and the population on longer time scales is described by an effective dynamics, in which early time transients feature quantum anti-Zeno behavior\cite{zeno3}. In this section we show explicitly that this transient dynamics results in violations of Fermi's Golden rule and detailed balance   of population kinetics. In section (\ref{sec:eomlr}) we obtain the equation of motion for a coherent (DAQ) condensate from the time evolution of the reduced density matrix \emph{without} invoking the Markov and rotating wave approximations yielding a time non-local equation of motion  that includes a self-energy.  In this section we provide an alternative derivation of the  equation of motion of the (DAQ) coherent condensate from quantum many body linear response, thereby establishing a direct bridge between the (QME) (without or with the Markov approximation) and quantum many body theory.  A multitime scale solution reveals the transient behavior including the dynamics of  quasiparticle formation.

 This equivalence allows us to provide a Feynman diagram-based interpretation of some of the approximations usually involved in the (QME), such as factorization. As a corollary of the linear response analysis we show that an emergent (DAQ) coherent condensate induces an expectation value of the Chern-Simons density, and relate the topological susceptibility to the (DAQ) self energy. In section (\ref{sec:discussion}) we provide a Feynman diagram-based interpretation of the origin of the (DAQ)-bath correlations, discuss important caveats and  draw more general lessons on the (QME) approach to non-equilibrium dynamics. Our conclusions and further questions are summarized in section (\ref{sec:conclusions}). Two appendices are devoted to the derivation of the correlation functions of the Chern-Simons density and relevant properties.

\section{Emergent dynamical axion quasiparticles:}\label{sec:daq}

The effective action for the emergent   dynamical axion field in topological insulators is given by\cite{xiao,rundong,jing,nomura,narang,gooth,gos,yu,mottola}\footnote{The vacuum dielectric constant and  vacuum permittivity  and a factor $4\pi$ have been  absorbed into a redefinition of the gauge fields and   the speed of light is set to $c=1$. }
\be  {S} = \int d^3x dt \Bigg\{ \frac{1}{2}\Big(  \vec{E}^2- {\vec{B}^2}  \Big)+\frac{\mathcal{J}}{2} \Bigg(\Big(\frac{\partial \delta \theta}{\partial t}\Big)^2- \Big(\vec{v}\cdot \vec{\nabla}\delta \theta\Big)^2-m^2\, \delta \theta^{\,^2} \Bigg) + \frac{\alpha}{ \pi} \,\delta \theta \,\vec{E}\cdot \vec{B}  \Bigg\}\,,\label{synac}  \ee where  the dynamical axion field $\delta \theta$ is related to the fluctuations of the Neel order parameter in the case of topological magnetic insulators, $\mathcal{J},\vec{v}$ are model and material dependent constants, $m$ is the dynamical axion mass\cite{rundong,jing,nomura}, and  $\alpha$ is the fine structure constant. We note that this effective action has been obtained by ``integrating out'' the electronic degrees of freedom and only describes the emergent collective   (DAQ) excitations interacting with electromagnetic fields \cite{xiao,rundong,jing,nomura,mottola}. As such, it has to be interpreted as an effective field theory valid below some energy scale at which the degrees of freedom that have been ``integrated out'' begin to influence the dynamics.  In 3D topological insulators this scale is determined by the bulk band gap $\simeq 10^{-2}-10^{-1}\,\mathrm{eV}$.

Redefining the (canonically normalized) dynamical axion field
\be \phi(\vx,t) = \sqrt{\mathcal{J}}\, \delta \theta(\vx,t) \,,\label{physint}\ee and assuming rotational invariance
the effective action for the   axionic quasiparticle field and the electromagnetic field is
\be {S}  = \int d^3x dt \Bigg\{ \frac{1}{2}\Big(  \vec{E}^2- {\vec{B}^2}  \Big)+\frac{1}{2} \Bigg(\Big(\frac{\partial \phi(\vx,t)}{\partial t}\Big)^2-  \Big(\vec{v}\cdot  \vec{\nabla}\phi(\vx,t)\Big)^2-m^2\, \phi^{\,2}(\vx,t)  \Bigg) + g \,\phi(\vx,t) \,\vec{E}\cdot \vec{B}  \Bigg\} \,,\label{synacfin} \ee with
$  g \equiv  {\alpha}/{\pi\sqrt{\mathcal{J}}}$.  While the form of the second bracket in $S$ may differ in the realizations of  dynamical axions in topological insulators and Weyl semimetals, the coupling to electromagnetism via the last, Chern-Simons  term is a   distinct hallmark of the coupling of   axions to $U(1)$ gauge fields,  a defining feature of axion electrodynamics\cite{wilczekaxion}.

 The time evolution   is determined by the quantum Hamiltonian operator. Since the interaction involves a time derivative, Hamiltonian quantization is
subtle  because the Chern-Simons interaction modifies the canonical momentum of the gauge field. The proper Hamiltonian quantization procedure  along with the interaction picture representation are  discussed in detail in reference\cite{boyhybrid}.    The main result is that  to leading order in the interaction the total Hamiltonian is
\be H = H_{0\gamma}+H_{0\phi}+H_i \,,\label{totHam}\ee with $H_{0\gamma},H_{0\phi}$ are the free field Hamiltonians for electromagnetism, and (DAQ) respectively and to leading order in the coupling $g$     the interaction Hamiltonian   is given by (see ref.\cite{boyhybrid})
\be H_i = - g\,\int d^3x \,  \phi(\vx)\,\mathcal{C}(\vx)   \,,\label{HIoft2}\ee where to consolidate notation we introduced the Chern-Simons density composite operator
\be \mathcal{C}(\vx,t) \equiv \vec{E}(\vx,t)\cdot\vec{B}(\vx,t)\,.\label{CSop}\ee

Free field quantization is achieved by expanding
 \be \phi(\vx,t) =  \sum_{\vk} \frac{1}{\sqrt{2VE_\phi(\vk)}}\,\Big[a(\vk)\,e^{-iE_\phi(\vk)t}\,e^{i\vk\cdot \vx} + a^\dagger(\vk)\,e^{iE_\phi(\vk)t}\,e^{-i\vk\cdot \vx}  \Big]\,,\label{daq}\ee where $V$ is the quantization volume and
 \be E_\phi(\vk) = \sqrt{(\vec{v}\cdot\vk)^2+m^2} \,,\label{energies} \ee are the single (DAQ) energies. In Coulomb gauge   the free field vector potential is given by
  \be \vec{A}(\vx,t) =  \sum_{\vk}\sum_{\lambda=1,2}  \frac{\hat{\vec{\epsilon}}_\lambda(\vk)}{\sqrt{2k V}}\, \Big[b_{\lambda}(\vk)\,e^{-ik t}\,e^{i\vk\cdot \vx} + b^\dagger_{\lambda}(\vk)\,e^{ikt}\,e^{-i\vk\cdot \vx}  \Big]\,,\label{vecpot2}\ee where the annihilation and creation operators obey the usual canonical commutation relations and the real linear polarization unit vectors $\hat{\vec{\epsilon}}_{\lambda}(\vec{k})$ are defined so that these along with the unit vector $\hat{\vec{k}}$ form a right handed triad (see appendix (\ref{app:ebcoup})). The free electric and magnetic fields are given by

\be \vec{E}(\vx,t) = - \frac{\partial}{\partial t}\vec{A}(\vx,t)~,~ \vec{B}(\vx,t) = \nabla \times A(\vx,t) \,. \label{ebfields}\ee
   In terms of these annihilation and creation operators which all obey canonical commutation relations,  the free field Hamiltonians  are

  \be H_{0\gamma} = \sum_{\vk}\sum_{\lambda=1,2}\, k\,\Big[b^\dagger_{\lambda}(\vk) b_\lambda(\vk)+ \frac{1}{2}\Big] ~~;~~ H_{0\phi} = \sum_{\vk} E_{\phi}(\vk) \, \Big[a^\dagger({\vk}) a({\vk})+\frac{1}{2}\Big] \,.\label{freehams}\ee
  The vacuum state for the (DAQ) is defined as
  \be a({\vk})\ket{0} =0 \,.\label{vacuum} \ee

  The observation of coherent oscillations   in two dimensional ($MnBi_2Te_4$)\cite{jianq} interpreted as a manifestation of (DAQ) motivates us to consider instead a coherent state  \be \ket{{\alpha}} = \Pi_{\vk}\, e^{- \frac{1}{2}|\alpha({\vk})|^2}\,e^{-\alpha({\vk}) a^\dagger({\vk})}\,\ket{0}\,, \label{cs} \ee it is an eigenstate of the annihilation operator,
\be a({\vk}) \ket{{\alpha}} = \alpha({\vk}) \ket{\bf{\alpha}} \,,  \label{eigenb}\ee  and describes a Poisson distribution of quanta of the free (DAQ) field.  The vacuum state is simply obtained by setting $\alpha({\vk}) =0\,\forall \vk$. The expectation values of the free (DAQ) field (\ref{daq})  and its canonical momentum $\pi(\vx,t) = \dot{\phi}(\vx,t)$ in this coherent state are
\bea \bra{\alpha} \phi(\vx,t) \ket{\alpha}  & = & \overline{\phi}(\vx,t) = \sum_{\vk}\frac{1}{\sqrt{2VE_{\phi}(\vk)}} \Big[\alpha({\vk})\,e^{-iE_\phi(\vk)t}   +\alpha^{*}({-\vk})\,e^{iE_\phi(\vk)t}    \Big]\, \,e^{i \vk\cdot \vx}\,,\label{expealp}\\
 \bra{\alpha} \pi(\vx,t) \ket{\alpha}  & = & \overline{\pi}(\vx,t) =  {-i}  \sum_{\vk}\sqrt{\frac{ E_{\phi}(\vk)}{ {2V }}} \Big[\alpha({\vk})\,e^{-iE_\phi(\vk)t}   -\alpha^{*}({-\vk})\,e^{iE_\phi(\vk)t}    \Big]\, \,e^{i \vk\cdot \vx}\,, \label{expepialp}
\eea
We refer to this expectation value as a free (DAQ) \emph{condensate}.

We consider the electromagnetic field as a bath in thermal equilibrium at a temperature $T=1/\beta$ described by the density matrix
\be \rho_{\gamma} = \frac{e^{-\beta H_{0\gamma}}}{\mathrm{Tr}\, e^{-\beta H_{0\gamma}}}\,,\label{emrho}\ee where for consistency, $T = 1/\beta$ is understood to be well below the scale that limits the validity of the Lagrangian (\ref{synacfin}) as an effective field theory so as not to excite the high energy degrees of freedom that have been integrated out, in fact in our study we consider $T$ to be of the same order as the single (DAQ) energy. The total density matrix at the initial time $t=0$ is assumed to
be
\be \rho(0) = \ket{\alpha}\bra{\alpha}\otimes \rho_{\gamma}\,.\label{inirho}\ee

This initial density matrix is off-diagonal in the basis of the occupation numbers of free (DAQ) quanta with the following expectation values which  will be needed in the analysis below
\bea && \mathrm{Tr}\, a^\dagger(\vq) \rho(0) = \alpha^*(\vq)~;~  \mathrm{Tr}\, a(\vq) \rho(0) = \alpha(\vq)~~;~~  \mathrm{Tr}\, a^\dagger(\vq)a^\dagger(-\vq) \rho(0) = \alpha^*(\vq)\alpha^*(-\vq)\nonumber \\ && \mathrm{Tr}\, a(\vq) a(-\vq)\rho(0) = \alpha(\vq)\alpha(-\vq)~~;~~N(\vq;0) =  \mathrm{Tr}\, a^\dagger(\vq)a(\vq) \rho(0)= |\alpha(\vq)|^2\,.\label{iniexp}\eea

The time evolution in the Schroedinger picture is given by
\be \rho(t) = e^{-iHt}\,\rho(0)\,e^{iHt}\,,\label{rhot}\ee however it is convenient to pass to the interaction picture
\be \rho_I(t) = U(t)\,\rho(0)\,U^{-1}(t) ~~;~~ U(t) = e^{iH_0t} e^{-iHt} \,\label{rhoip}\ee where the unitary time evolution operator in the interaction picture $U(t)$ obeys
\be \frac{d}{dt} U(t) = -iH_I(t) U(t)~~;~~ U(0) =1 \,\label{timeevolU}\ee and
\be H_I(t) = e^{iH_0t}\, H_i \, e^{-iH_0t}\,\label{HIt} \ee  where $H_i$ is given by eqns. (\ref{HIoft2},\ref{CSop}).  $U(t)$ can be written as a  perturbative time-ordered Dyson series as usual,
\be U(t) = 1 + i\,g\, \int^t_0 \int d^3 x \, \phi_I(\vx,t')\,\mathcal{C}_I(\vx,t')\,dt' + \cdots\label{Upert}\ee

Operators in the interaction picture feature free field time evolution, namely
\be \mathcal{O}_I(t) = e^{iH_0t}\,\mathcal{O}(0) \, e^{-iH_0t} \,, \label{Aop}\ee and Heisenberg operators evolve in time as
\be \mathcal{O}(t) = e^{iHt} \mathcal{O}(0) e^{-iHt} = U^{-1}(t) \mathcal{O}_I(t) \,U(t) \,.\label{intheis}\ee relating Heisenberg to interaction picture operators.

In particular the interaction picture (DAQ) and electromagnetic fields are given respectively by free field expressions (\ref{daq},\ref{vecpot2}), namely
\bea
&&  \phi_I(\vx,t) =  \sum_{\vk} \frac{1}{\sqrt{2VE_\phi(\vk)}}\,\Big[a(\vk)\,e^{-iE_\phi(\vk)t}\,e^{i\vk\cdot \vx} + a^\dagger(\vk)\,e^{iE_\phi(\vk)t}\,e^{-i\vk\cdot \vx}  \Big]\\\label{daqip}
&& \vec{A}_I(\vx,t) =  \sum_{\vk}\sum_{\lambda=1,2}  \frac{\hat{\vec{\epsilon}}_\lambda(\vk)}{\sqrt{2k V}}\, \Big[b_{\lambda}(\vk)\,e^{-ik t}\,e^{i\vk\cdot \vx} + b^\dagger_{\lambda}(\vk)\,e^{ikt}\,e^{-i\vk\cdot \vx}  \Big] \,,
\label{emip} \eea and the operators $a^\dagger(\vk),a(\vk)~;~b^\dagger_\lambda(\vk),b_\lambda(\vk)$ do not depend on time, an important aspect that will be used below to obtain the equation of motion of (DAQ) condensates. We have appended the subscript $I$ denoting interaction picture fields to emphasize that these are in the interaction picture with free field time evolution,  to  distinguish them from the Heisenberg  fields in the fully interacting theory, which evolve in time with the total Hamiltonian including the interaction. This is   an important distinction that will be highlighted below within the context of linear response.

The expectation value of Heisenberg field operators  $\mathcal{O}(\vx,t)$ in the initial density matrix $\rho(0)$ are related to expectation values of the interaction picture operators as
\be \langle \mathcal{O}(\vx,t) \rangle \equiv \mathrm{Tr} \, \mathcal{O}(\vx,t)\,\rho(0) = \mathrm{Tr} \, \mathcal{O}_I(\vx,t)\,\rho_I(t)\,.\label{Oex}\ee

 The interaction picture density matrix $\rho_I(t)$ obeys the time evolution equation
\be \dot{\rho}_I(t) = -i [H_I(t),\rho_I(t)] \,,\label{rhodot}\ee with formal solution
\be \rho_I(t) = \rho_I(0)-i\,\int^t_0 [H_I(t'),\rho_I(t')] \,dt' \,. \label{solu}\ee This solution is inserted back into (\ref{rhodot}) leading to the iterative equation
\be \dot{\rho}_I(t) = -i[H_I(t),\rho_I(0)]- \int^t_0[H_I(t),[H_I(t'),\rho_I(t')]]\,dt' \,.\label{itereq}\ee This equation cannot be solved exactly and several approximations are usually invoked\cite{breuer,linme}.

\vspace{1mm}

\textbf{a) Factorization: } the total density matrix is assumed to factorize at all times as
\be \rho_I(t) = \rho_{I\phi}(t)\otimes \rho_{\gamma}  \,.\label{fact}\ee As argued in reference\cite{linme}, although we assumed factorization at the initial time, it is expected that the time evolution generates correlations between the (DAQ) and the bath at order $g^2$, suggesting that
\be \rho_I(t) = \rho_{I\phi}(t)\otimes \rho_{\gamma}+ g^2\,\rho_{corr}(t) \,,\label{corre}\ee the correlation term is in general non-factorizable. It is  of order $g^2$ and would generate contributions of order
$g^4$ to the equation (\ref{itereq}) since $H_I \propto g$, therefore for weak coupling and up to order $g^2$ we neglect the correlation term in (\ref{corre}). In principle even with a factorized form, the density matrix for the bath is expected to develop time evolution, however under the assumptions that the bath has a large number of degrees of freedom and weak coupling, it is assumed to remain in thermal equilibrium\cite{breuer,linme}. Both of these aspects will be discussed further in section (\ref{sec:discussion}) where we provide a quantum many body theory, Feynman diagram-based interpretation of these assumptions.

The \emph{reduced density matrix} for the (DAQ) field $\phi(\vx,t)$ is obtained by taking the trace of the full density matrix $\rho_I(t)$ in the Hilbert space of the bath ($\gamma$) degrees of freedom,  namely
\be \rho_{I\phi}(t) = \underset{\{\gamma\}}{\text{Tr}} \,\rho_I(t) \,.\label{rhophi}\ee  Upon taking the trace over the electromagnetic (bath) fields, the first term in (\ref{itereq}) vanishes since
\be Tr\,\vec{E}(\vx,t)\cdot \vec{B}(\vx,t)\,\rho_\gamma =0 \,,\label{vani}\ee because $\vec{E}\cdot\vec{B}$ is odd under both parity and time reversal, whereas the equilibrium density matrix of electromagnetic fields is   even. As a result, the time evolution of the reduced density matrix is given by

\bea \dot{\rho}_{I\phi}(t) & =  & -g^2\int^t_0 dt' \int d^3x \int d^3 x'\Bigg\{G^>(\vx-\vx',t-t')\,\Big[ \phi_I(\vx,t) \, \phi_I(\vx^{\,'},t')\, {\rho}_{I\phi}(t')-  \phi_I(\vx^{\,'},t') \,  {\rho}_{I\phi}(t')\, \phi_I(\vx,t)\Big] \nonumber \\ & + & G^<(\vx-\vx',t-t')\,\Big[ {\rho}_{I\phi}(t') \,\phi_I(\vx^{\,'},t')\, \phi_I(\vx,t)- \phi_I(\vx,t) \, {\rho}_{I\phi}(t') \, \phi_I(\vx^{\,'},t')  \Big] \Bigg\}\,,\label{Linblad}\eea

where $\phi_I(\vx,t)$ is the (DAQ) field in interaction picture (\ref{daq}), and the correlation functions
\bea  G^>(\vx-\vx',t-t') & = & \underset{\{\gamma\}}{\text{Tr}} \,\rho_\gamma \, \mathcal{C}_I(\vx,t) \, \mathcal{C}_I(\vx^{\,'},t') \label{ggreat} \\
G^<(\vx-\vx',t-t') & = &  \underset{\{\gamma\}}{\text{Tr}} \, \rho_\gamma\, \mathcal{C}_I(\vx^{\,'},t')\, \mathcal{C}_I(\vx,t)  \,,  \label{gless} \eea where the Chern-Simons composite operator $\mathcal{C}_I(\vx,t)$  is given by equation (\ref{CSop}) with the electromagnetic field  in the interaction picture (\ref{emip}).  It is clear from (\ref{Linblad}) that the time evolution of the reduced density matrix is trace preserving.

For any interaction picture operator $\mathcal{O}_{I\phi}(\vx,t)$ that acts solely on the (DAQ) degrees of freedom, neglecting the correlation term in (\ref{corre})  its expectation value in the  full density matrix $\rho_I(t)$   is given by
\be \langle \mathcal{O}_{I\phi}(\vx,t) \rangle \equiv \mathrm{Tr}\,\mathcal{O}_{I\phi}(\vx,t)\,\rho_{I}(t) = \mathrm{Tr}\,\mathcal{O}_{I\phi}(\vx,t)\,\rho_{I\phi}(t) \,,\label{aveO}\ee consequently
\be \frac{d}{dt} \langle \mathcal{O}_{I\phi}(\vx,t) \rangle = \langle \dot{\mathcal{O}}_{I\phi}(\vx,t) \rangle + \mathrm{Tr}\,\mathcal{O}_{I\phi}(\vx,t)\,\dot{\rho}_{I\phi}(t) \,,\label{timeder}  \ee where the dot stands for the time derivative.

\vspace{1mm}

\textbf{b) (Partial) Markov approximation:}
 the second approximation entails replacing $\rho_{I\phi}(t') \rightarrow \rho_{Ia}(t)$ in the time integral. This is usually referred to as a Markov approximation and   is  justified in weak coupling, as can be seen by   considering the first term   under the integrals in (\ref{Linblad}) as an example. It can be written as
\be -g^2 \, \phi_I(\vx,t) \int^t_0 \frac{d\mathcal{H}(t')}{dt'} \, {\rho}_{I\phi}(t') \,dt' ~~;~~  \mathcal{H}(t') \equiv \int^{t'}_0 \phi_I({\vx}',t'') \, G^>(\vx-{\vx}',t-t'')dt'' \label{incha}\ee which upon integration by parts yields
\be -g^2 \,\phi_I(\vx,t)  \mathcal{H}(t) \hat{\rho}_{I\phi}(t) + g^2 \,  \phi_I(\vx,t) \int^t_0  \mathcal{H}(t') \,\frac{d{\rho}_{I\phi}(t')}{dt'} dt' \label{incha2}\ee in the second term $d{\rho}_{I\phi}(t')/dt' \propto g^2$ so this   term yields a contribution that is formally of order $g^4$ and will be neglected consistently with keeping only up to second order contributions. The same analysis is applied to all the other terms in (\ref{Linblad}) with the conclusion that in weak coupling and  to leading order $(g^2)$  the Markovian approximation  ${\rho}_{I\phi}(t') \rightarrow \hat{\rho}_{I\phi}(t)$ is justified.

Therefore in the ``partial'' Markov approximation the quantum master equation becomes

\bea \dot{\rho}_{I\phi}(t) & =  & -g^2\int^t_0 dt' \int d^3x \int d^3 x'\Bigg\{G^>(\vx-\vx^{\,'},t-t')\,\Big[ \phi_I(\vx,t) \, \phi_I(\vx^{\,'},t')\, {\rho}_{I\phi}(t)-  \phi_I(\vx^{\,'},t') \,  {\rho}_{I\phi}(t)\, \phi_I(\vx,t)\Big] \nonumber \\ & + & G^<(\vx-\vx^{\,'},t-t')\,\Big[ {\rho}_{I\phi}(t) \,\phi_I(\vx^{\,'},t')\, \phi_I(\vx,t)- \phi_I(\vx,t) \, {\rho}_{I\phi}(t) \, \phi_I(\vx^{\,'},t')  \Big] \Bigg\}\,.\label{Linbladmarkov}\eea

In the literature, the Markov approximation is taken further by extending the upper limit of the time integrals $t\rightarrow \infty$ under the assumption of a wide separation of time scales\cite{breuer,linme,nonmarkov,vega}. However, we will \emph{not} implement this
extra step, keeping the finite time integral to explore the consequences of transient dynamics.

The correlation functions $G^{>}(x-x'),G^<(x-x')$ are obtained in appendix (\ref{app:corre}) as   Lehmann-spectral  representations\cite{fetter,mahan}, they are given by
\bea  G^>(x-x') & = & \int \frac{d^3q}{(2\pi)^3} \int \frac{dq_0}{2\pi} ~\rho^>(q_0,\vq)\,e^{-iq_0(t-t')}~e^{i \vq\cdot(\vx-\vx')} \label{ggreatspec}\\
G^<(x-x') & = & \int \frac{d^3q}{(2\pi)^3} \int \frac{dq_0}{2\pi} ~\rho^<(q_0,\vq)\,e^{-iq_0(t-t')}~e^{i \vq\cdot(\vx-\vx')} \,,\label{glessspec}\eea where the spectral densities obey the relation
\be \rho^>(-q_0,\vq)= \rho^<(q_0,\vq)\,,\label{rela3}  \ee and fulfill the Kubo-Martin-Schwinger condition\cite{kms}
\be \rho^<(q_0,\vq) = e^{-\beta\,q_0} \,\rho^>(q_0,\vq)\,, \label{kms} \ee
which  is  a consequence of   the electromagnetic fields   being  in thermal equilibrium. Introducing the spectral density
\be \rho(q_0, \vq) = \rho^>(q_0,\vq)-\rho^<(q_0,\vq)\,,  \label{specdens}\ee the Kubo-Martin-Schwinger condition (\ref{kms}) leads to the following relations
\bea \rho^>(q_0,\vq) & = &  [1+n(q_0)]\,\rho(q_0,\vq) \label{rhogreat2} \\\rho^<(q_0,\vq) & = &   n(q_0)\,\rho(q_0,\vq) \label{rholess2} \eea where $n(q_0) = [e^{\beta\,q_0}-1]^{-1}$ is the Bose-Einstein distribution function at temperature $T=1/\beta$.  The above relations are proven in appendix (\ref{app:corre}), they are general,   and rely only on that the electromagnetic field, namely the bath,  is in thermal equilibrium. The spectral density $\rho(q_0,\vq)$ is obtained in detail in appendix (\ref{app:ebcoup}).

{\textbf{c:) Rotating wave approximation\cite{breuer}-\cite{linme}:}}  in writing the products $\phi_I(\vx,t)~\phi_I(\vx^{\,'},t')$ of interaction picture field operators (\ref{daqip}) in (\ref{Linbladmarkov}) there are two types of terms  with very different time evolution. Terms of the form
\be a^\dagger(\vq)~a(\vq)~e^{  i{ E_{\phi}(\vq)(t-t')}}\,, \label{ada} \ee   are ``slow'', and terms of the form
\be a^\dagger(\vq)~a^\dagger(-\vq)~ e^{2iE_{\phi}(\vq) t'}~e^{i{E_{\phi}(\vq)(t-t')}}~~;~~ a(\vq)~a(-\vq)~ e^{-2iE_{\phi}(\vq) t'}~e^{-i{E_{\phi}(\vq)(t-t')}}\,, \label{nonrot}\ee are fast, the extra rapidly varying phases $e^{ \pm 2iE_{\phi}(\vq) t'}$ lead to rapid dephasing on time scales $\simeq 1/E_{\phi}(\vq)$ and do not yield resonant (nearly energy conserving) contributions. Neglecting these terms is tantamount to neglecting non-resonant terms that average out over   time scales  $\simeq 1/E_{\phi}(\vq)$.   These terms only give perturbatively small transient contributions and are discussed in section (\ref{sec:discussion}). Keeping only the slow terms  and neglecting the   terms (\ref{nonrot})  defines the ``rotating wave approximation'' ubiquitous in quantum optics\cite{breuer,zoeller}.

Implementing the  Markov  approximation ${\rho}_{I\phi}(t')\rightarrow {\rho}_{I\phi}(t)$, and the    rotating wave  approximation keeping only terms of the form $a^\dagger a, a a^\dagger$,  using the spectral representation of the correlators (\ref{ggreatspec},\ref{glessspec})    and carrying out the spatial and temporal integrals we finally obtain the  \emph{Lindblad} form\cite{breuer,zoeller,lin,gori,pearle,linme} of the quantum master equation (QME)
\bea \dot{{\rho}}_{I\phi}(t)  & = &  \sum_{\vq} \Bigg\{   -i\,\Delta(\vq,t)~\Big[a^\dagger(\vq)~a(\vq), {\rho}_{I\phi}(t) \Big] \nonumber \\
& - & \frac{\Gamma^>(\vq,t)}{2} \Big[a^\dagger(\vq)~a(\vq)~ {\rho}_{I\phi}(t) + {\rho}_{I\phi}(t)~a^\dagger(\vq)~a(\vq) - 2 \, a(\vq)~{\rho}_{I\phi}(t)~a^\dagger(\vq) \Big]\nonumber \\
& - & \frac{\Gamma^<(\vq,t)}{2} \Big[a(\vq)\,a^\dagger(\vq)~ {\rho}_{I\phi}(t) + {\rho}_{I\phi}(t)~a(\vq)\,a^\dagger(\vq) - 2 \, a^\dagger(\vq)~{\rho}_{I\phi}(t)~a(\vq) \Big] \Bigg\} \,,
\label{Linfin} \eea  where
\be \Delta(\vq,t) = \frac{g^2}{2\,E_{\phi}(\vq)} \,\int \frac{dq_0}{2\pi} \,\rho(q_0,q)\,\frac{\Big[1-\cos[(E_{\phi}(\vq)-q_0)t] \Big]}{(E_{\phi}(\vq)-q_0)}\,, \label{Roftim}\ee
\be \Gamma^>(\vq,t) = \frac{g^2}{ E_{\phi}(\vq)}  \,\int \frac{dq_0}{2\pi} \, \rho(q_0,q)\,\big[1+n(q_0)\big]\frac{ \sin[(E_{\phi}(\vq)-q_0)t] }{ (E_{\phi}(\vq)-q_0)} \,, \label{gamgre}\ee
\be \Gamma^<(\vq,t) = \frac{g^2}{E_{\phi}(\vq)}  \,\int   \, \rho(q_0,q)\, n(q_0) \frac{ \sin[(E_{\phi}(\vq)-q_0)t] }{(E_{\phi}(\vq)-q_0)}\, \frac{dq_0}{2\pi} \,, \label{gamles}\ee where $n(q_0) = [e^{\beta q_0}-1]^{-1}$,
and we introduce
\be \Gamma(\vq,t) = \Gamma^>(\vq,t)-\Gamma^<(\vq,t)= \frac{g^2}{E_{\phi}(\vq)} \,\int  \,\rho(q_0,q)\,\frac{ \sin[(E_{\phi}(\vq)-q_0)t] }{(E_{\phi}(\vq)-q_0)}\, \frac{dq_0}{2\pi} \,. \label{gamadif}\ee

The second and third lines in (\ref{Linfin}) are called the \emph{dissipator}\cite{breuer}, these are non-Hamiltonian, purely dissipative terms, however it follows from the (QME) (\ref{Linfin}) that the trace of the reduced density matrix is conserved. It is argued in refs. \cite{lin,gori,pearle}   that the equation (\ref{Linfin}) is the most general linear evolution equation that preserves unit trace and  Hermiticity of the density matrix.

In the usual Born-Markov approximation the coefficient functions in the Lindblad (QME), $\Delta(\vq,t),\Gamma^{\lessgtr}(\vq,t)$ are evaluated at $t =\infty$\cite{breuer,linme,nonmarkov,vega}, instead, we keep the full time dependence of these functions, in particular $\Gamma^{\lessgtr}(\vq,t)$ to study transient phenomena associated with the build-up of the quasiparticle.

From the relation (\ref{Oex}) the expectation value of the \emph{interacting} (DAQ) field $\phi(\vx,t)$ is given by the expectation value of the field in interaction picture $\phi_I(\vx,t)$ in the interaction picture reduced density matrix $\rho_{I\phi}$, namely the interacting \emph{condensate} is given by

\be \langle \phi(\vx,t) \rangle \equiv \overline{\phi}(\vx,t)= \mathrm{Tr}\, \phi_I(\vx,t) {\rho}_{I\phi}(t) = \sum_{\vq} \frac{1}{\sqrt{2\,V\,E_{\phi}(\vq)}}\Big[  \langle a({\vq}) \rangle(t)\,e^{-iE_{\phi}(\vq) t} +  \langle a^\dagger({-\vq})\rangle(t)  \, e^{iE_{\phi}(\vq) t} \Big]\,   e^{i\vq\cdot\vx}\,, \label{expvalI}  \ee where
\be  \langle a({\vq}) \rangle(t)=\underset{\{\phi\}}{\text{Tr}}\Big(a({\vq})\,  {\rho}_{I\phi}(t) \Big) ~~ ; ~~\langle a^\dagger({-\vq})\rangle(t) = \underset{\{\phi\}}{\text{Tr}}\Big(a^\dagger({-\vq})\, {\rho}_{I\phi}(t) \Big) \,. \label{aves} \ee   Using equations (\ref{aveO},\ref{timeder}) and that the  operators $a(\vq),a^\dagger(\vq)$  do not depend on time in the interaction picture, we find
\bea \frac{d}{dt} \langle a({\vq})\rangle(t)  & = &  \Big[-i\,\Delta(\vq,t)-\frac{\Gamma(\vq,t)}{2} \Big]\, \langle a({\vq}) \rangle(t) \nonumber \\\frac{d}{dt}\langle a^\dagger({\vq})\rangle(t)   & = &  \Big[i\,\Delta(\vq,t)-\frac{\Gamma(\vq,t)}{2} \Big]\, \langle a^\dagger({-\vq})\rangle(t) \,, \label{aveaad}\eea
with solutions
\be  \langle a({\vq})\rangle(t) = \Big[e^{-i \int^t_0 \Delta(\vq,t')\,dt'}\,e^{-\frac{1}{2}\int^t_0 \Gamma(\vq,t')\,dt'}  \Big] \,\alpha(\vq)\,, \label{aoftsol}\ee and the hermitian conjugate for $\langle a^\dagger({\vq})\rangle(t)$. Similarly
for the bilinears
\bea \frac{d}{dt} \langle a_{\vq}~ a_{-\vq} \rangle (t) & = &  \Big[-2i\,\Delta(\vq,t)  - \Gamma(\vq,t)\Big] \langle a_{\vq}~ a_{-\vq} \rangle (t) \nonumber \\
 \frac{d}{dt} \langle a^\dagger_{\vq}~ a^\dagger_{-\vq} \rangle(t)  & = &  \Big[2i\,\Delta(\vq,t)  - \Gamma(\vq,t)\Big] \langle a^\dagger_{\vq} ~ a^\dagger_{-\vq} \rangle(t)\,,  \label{bilins}\eea with solution
 \be  \langle a_{\vq}~ a_{-\vq} \rangle (t) =  \Big[e^{-2i \int^t_0 \Delta(\vq,t')\,dt'}\,e^{- \int^t_0 \Gamma(\vq,t')\,dt'}  \Big] \alpha(\vq)\alpha(-\vq) \,, \label{aaoftsol}  \ee
and the hermitian conjugate for $\langle a^\dagger_{\vq}~ a^\dagger_{-\vq} \rangle(t)$. We refer to (\ref{aoftsol},\ref{bilins}) and their hermitian conjugate as \emph{coherences} their non-vanishing imply that the density matrix is off-diagonal in the occupation number basis.  From the evolution equations (\ref{aveaad},\ref{bilins}) it is clear that $\Delta(\vq,t)$ is a time dependent renormalization of the (DAQ) single particle energy $E_{\phi}(\vq)$, and $\Gamma(\vq,t)$ an effective time dependent decay rate. The solutions (\ref{aoftsol},\ref{aaoftsol}) and their hermitian conjugates  indicate \emph{decoherence} in the occupation number basis, since the coherences decay $\propto e^{- \int^t_0 \Gamma(\vq,t')\,dt'}$.

The expectation value of the number operator
\be N(\vq,t) = \underset{\{\phi\}}{\text{Tr}}\, {\rho}_{I\phi}(t)\, a^\dagger({\vq}) \, a({\vq})\,\label{numop}\ee is found to obey the quantum kinetic equation

\be \frac{dN(\vq,t)}{dt} =  \underset{\{\phi\}}{\text{Tr}}\Big\{ a^\dagger({\vq}) \, a({\vq}) ~\dot{{\rho}}_{I\phi}(t)\Big\} = -\Gamma(\vq,t) N(\vq,t) + \Gamma^<(\vq,t) \,, \label{qke}\ee which  can be written  as
 \be \frac{dN(\vq,t)}{dt} =  \Gamma^<(\vq,t)\,\Big(1+N(\vq,t)\Big) - \Gamma^>(\vq,t)\, N(\vq,t)  \,.\label{qkebe}  \ee
 In this form the quantum kinetic equation   has a simple   interpretation as the Boltzmann equation for the distribution function $N(\vq,t)$ of the form $dN/dt = (\mathrm{gain})- (\mathrm{loss})$  with a loss  term $\Gamma^>(\vq,t)\, N(\vq,t)$ arising from the decay of a (DAQ) into two photons $\phi \rightarrow 2 \gamma$ and a gain term $\Gamma^<(\vq,t)\,\Big(1+N(\vq,t)\Big)$ from recombination $2\gamma \rightarrow \phi$
 in the medium in which photons remain in thermal equilibrium, which is one of the main assumptions in the derivation of the quantum master equation in the Born-Markov approximation and explicitly stated in the factorized form (\ref{fact}).

 The time dependent distribution function of (DAQ) is the solution of the quantum kinetic equation (\ref{qke}) which is given by
   \be  N(\vq,t) = e^{-\int^t_0 \Gamma(\vq,t')\,dt'}\,\Big[|\alpha(\vq)|^2 + \int^t_0 \Gamma^<(\vq,t')\,e^{\int^{t'}_0 \Gamma(\vq,t'')\,dt''}\,dt' \Big]\,.\label{disfun} \ee

If, as is customary  in Fermi's golden rule,  the long time limit is taken
 \be  \frac{ \sin[(E_{\phi}(\vq)-q_0)t] }{(E_{\phi}(\vq)-q_0)} ~~{}_{\overrightarrow{t\rightarrow \infty}} ~~\pi \,\delta(q_0-E_\phi(\vq))\, , \label{sinlt}\ee   it follows that
 \bea   & \Gamma^<(\vq,t) & ~~{}_{\overrightarrow{t\rightarrow \infty}} ~~ \equiv \Gamma^<(\vq,\infty) = \Gamma(\vq,\infty)\,n(E_\phi(\vq))\,\label{glessifty}  \\
 & \Gamma^>(\vq,t) & ~~{}_{\overrightarrow{t\rightarrow \infty}} ~~ \equiv \Gamma^>(\vq,\infty) = \Gamma(\vq,\infty)\,\big(1+n(E_\phi(\vq))\big)\,, \label{ggreatifty}\eea where $n(E_\phi(\vq))$ is the Bose-Einstein distribution function for (DAQ) in thermal equilibrium and
  \be   \Gamma(\vq,\infty)=\frac{g^2}{2 E_{\phi}(\vq)}\,\rho(E_{\phi}(\vq),q)\,,  \label{gamafgr} \ee is the decay rate in agreement with Fermi's golden rule.
  Therefore, in the long time limit the ratio of the forward (decay-loss) and backward (recombination-gain) rates obeys the detailed balance relation
  \be \frac{\Gamma^>(\vq,\infty)}{\Gamma^<(\vq,\infty)} = \frac{1+n(E_\phi(\vq))}{n(E_\phi(\vq))}\,.\label{detbal}\ee  Replacing the rates by their long time limits (\ref{glessifty},\ref{ggreatifty}) into the quantum kinetic equation (\ref{qkebe}) yields as solution
  \be N(\vq,t) = n(E_\phi(\vq))+ \Big(N(\vq,0)-n(E_\phi(\vq))\Big) \,e^{-\Gamma(\vq,\infty)\,t} \,,\label{apptoeq}\ee which is the usual dynamics of approach to thermal equilibration with the bath on a time scale $\simeq 1/\Gamma(\vq,\infty)$ and a direct consequence of the detailed balance relation (\ref{detbal}) for constant gain and loss terms which ensures that the equilibrium distribution function is a fixed point of the quantum kinetic equation (\ref{qkebe}).

  However, taking the (infinite) long time limits in the rates \emph{before} solving the equation for the time evolution of the population not only assumes a clean separation
  of time scales but also ignores transient phenomena that describes the formation of the quasiparticle, and violations Fermi's Golden Rule and of detailed balance.

  \section{Transient dynamics: a quasiparticle is born}\label{sec:transient}

The solutions of     equations (\ref{aveaad},\ref{bilins}, \ref{qke}) are straightforward in terms of time integrals of $\Delta(\vq,t);\Gamma(\vq,t)$ whose  long time limits are given by

 \be \int^t_0 \Delta(\vq,t') dt'  =  t\,\frac{g^2}{2\,E_{\phi}(\vq)} \int^\infty_{-\infty}   \frac{\rho(q_0,\vq)}{(E_{\phi}(\vq)-q_0)} \,\Bigg[ 1-\frac{\sin(E_{\phi}(\vq)-q_0)\,t}{(E_{\phi}(\vq)-q_0)\,t} \Bigg]\,\frac{dq_0}{(2\pi)} ~~{}_{\overrightarrow{t\rightarrow \infty}} ~~ t\,\delta E_{\phi}(\vq)  \label{realpartofE}  \ee where
  \be \delta E_{\phi}(\vq) = \frac{g^2}{2\,E_{\phi}(\vq)}\, \int^\infty_{-\infty} \mathcal{P}\Bigg[ \frac{\rho(q_0,q)}{(E_{\phi}(\vq)-q_0)} \Bigg]\,\frac{dq_0}{2\pi}\,,\label{renfreq}\ee is a renormalization of the single particle energy  $E_{\phi}(\vq)$ and  $\mathcal{P}$ stands for the principal part, and
  \bea  \int^t_0 \Gamma(\vq,t')\,dt'  & = &   \frac{g^2}{ E_{\phi}(\vq)}\, \int_{-\infty}^{\infty} \frac{dq_0}{2\pi} \,  {\rho(q_0,q)}\,\Big[\frac{ 1-\cos\big[(q_0-E_{\phi}(\vq))t\big]}{( q_0-E_{\phi}(\vq))^2} \Big]\nonumber \\ & {}_{\overrightarrow{t\rightarrow \infty}}& \Gamma(\vq,\infty) \, t +  \frac{g^2}{ E_{\phi}(\vq)}\,  \int_{-\infty}^{\infty} \frac{dq_0}{2\pi} \,\mathcal{P}  \frac{\rho(q_0,q)}{( E_{\phi}(\vq)-q_0)^2}\,, \label{gamasy}\eea where $\Gamma(\vq,\infty)$  is given by (\ref{gamafgr}).  This long time limit is understood as follows: at long time, the region of $q_0$ in the integral around $E_\phi(\vq)$ of width $\Delta q_0 \simeq 2\pi/ t$ yields a contribution to the integral that grows $\propto \rho(E_\phi)\,t$, this is familiar from Fermi's golden rule. The oscillating cosine terms integrated in the remaining domain of $q_0$  namely   $|q_0-E_{\phi}(\vq)| > 2\pi/t$ average out by Riemann-Lebesgue, leaving solely the contributions that are not oscillatory (the ``1'' inside the bracket in equation (\ref{gamasy})) in the long time   limit after ``excising'' the region $|q_0-E_\phi(\vq)| \lesssim 2\pi/t$, yielding the principal value prescription, hence equation (\ref{gamasy}) follows.

   It is important to highlight that in the strict infinite time limit only the term that grows linearly with time remains in (\ref{gamasy}), thereby neglecting the constant term, this is valid at times long enough that the constant term can be effectively neglected. However, for the (DAQ) case $\rho(q_0,q) \simeq q^4_0$ for large $q_0$ and the integral in (\ref{gamasy}) diverges as $\Lambda^2_{max}$ with $\Lambda_{max}$ and upper frequency (ultraviolet) cutoff.   As   discussed below
   explicitly, this term is directly associated with the quasiparticle weight or ``residue'' (wave function renormalization) and the dressing of the (bare) particle into the quasiparticle  by quantum fluctuations.

  In the long time limit, and using the results (\ref{realpartofE},\ref{gamasy}) the solution of eqns. (\ref{aveaad},\ref{bilins}) are
  \be \langle a_{\vq} \rangle(t) = {Z}(\vq)\,  e^{-i\delta E_{\phi}(\vq)\, t}\,e^{-\frac{\Gamma(\vq,\infty)}{2}t } \,  \alpha({\vq})  ~~;~~ \langle a^\dagger({\vq}) \rangle(t) = {Z}(\vq)\,  e^{ i\delta E_{\phi}(\vq)\, t}\,e^{-\frac{\Gamma(\vq,\infty)}{2}t } \,\alpha^*(\vq)\,,\label{solubs}  \ee
   where we used the initial expectation values (\ref{iniexp}), and  to leading order in the coupling,
  \be Z(\vq) = 1- \frac{g^2}{ 2E_{\phi}(\vq)}\,  \int_{-\infty}^{\infty} \frac{dq_0}{2\pi} \,\mathcal{P}  \frac{\rho(q_0,q)}{( E_{\phi}(\vq)-q_0)^2} \label{wfun} \ee is the quasiparticle weight or ``residue'' (wave function renormalization).

  Inserting these results into the expression for the condensate (\ref{expvalI}) yields for its long time limit

  \be  \overline{\phi}(\vx,t)=  \sum_{\vq} \frac{Z(\vq)}{\sqrt{2\,V\,E_{\phi}(\vq)}}\Big[ \alpha(\vq)\,e^{-iE_{R\phi}(\vq) \, t}\,e^{-\frac{\Gamma(\vq,\infty)}{2}\,t} +  \alpha(-\vq)\,e^{iE_{R\phi}(\vq)\, t}\,e^{-\frac{\Gamma(\vq,\infty)}{2}\,t} \Big]\,   e^{i\vq\cdot\vx}\,, \label{expvalqme}  \ee
  where
  \be E_{R\phi}(\vq) = E_{\phi}(\vq)+\delta E_{\phi}(\vq)\,,\label{Eren}\ee is the renormalized single (DAQ) energy. Similarly for the bilinear coherences

  \bea && \langle a({\vq})\,a({-\vq}) \rangle(t) = {Z}^2(\vq)\,  e^{-2i\delta E_{\phi}(\vq) t}\,e^{- \Gamma(\vq,\infty)\, t } \,\alpha(\vq)\,\alpha(-\vq)\nonumber \\ &&  \langle a^\dagger({\vq}) a^\dagger({-\vq})\rangle(t) = {Z}^2(\vq)\,  e^{ 2i\delta E_{\phi}(\vq) t}\,e^{- \Gamma(\vq,\infty)\, t } \,\alpha^{*}(\vq)\,\alpha^{*}(-\vq) \,,\label{solubilins}  \eea
 A quasiparticle is generically characterized by its (renormalized) energy, its damping or decay rate and the quasiparticle weight or ``residue''. In the next section we establish a direct relation between these results obtained with the (QME) and the quantum  many body approach of obtaining the (renormalized) frequencies, damping or decay rates and quasiparticle weight (residue) from  the complex poles of single particle  Green's function including self-energy corrections.

The integral of the time dependent rate, equation (\ref{gamasy}) describes important transient dynamics, let us define $\int^t_0 \Gamma(\vq,t')\,dt' \equiv  \frac{g^2}{ E_{\phi}(\vq)}\, I(t)$,  where
\be I(t) = \int^\infty_0 \rho(q_0,\vq)\,F[q_0,\vq;t] \, \frac{dq_0}{2\pi} \,,\label{Ioft}\ee with
\be F[q_0,\vq;t] = \Big[\frac{ 1-\cos\big[(q_0-E_{\phi}(\vq))t\big]}{( q_0-E_{\phi}(\vq))^2} \Big]-\Big[\frac{ 1-\cos\big[(q_0+E_{\phi}(\vq))t\big]}{( q_0+E_{\phi}(\vq))^2} \Big] \equiv F_+[q_0,\vq;t]-F_{-}[q_0,\vq;t] \,,\label{fot}\ee where $F_{\pm}[q_0,\vq,t]$ feature singular denominators at $\pm E_{\phi}(\vq)$ and we used the property $\rho(-q_0,\vq)= - \rho(q_0,\vq)$ (see equation (\ref{rhofi})). The singular denominator in $F_+[q_0,\vq,t]$ within the domain of integration at $q_0 = E_\phi(\vq)$   makes this function to be strongly peaked near this value at large time, this is the familiar time dependent function of Fermi's golden rule, whose integral yields a contribution linear in time. It is convenient to separate the contribution of the region in $q_0$ that contains the singular denominator by writing
\be I(t) = I_p(t) + I_c(t) \,,\label{Isplit}\ee with
\bea I_p(t) & = &  \int^{\Lambda_p}_0 \rho(q_0,\vq)\,F[q_0,\vq;t] \, \frac{dq_0}{2\pi}~~;~~ \Lambda_p > E_\phi(\vq)\,\label{Ipin}\\
 I_c(t) & = &  \int^{\Lambda_{max}}_{\Lambda_p} \rho(q_0,\vq)\,F[q_0,\vq;t] \, \frac{dq_0}{2\pi}  \,.\label{Icin}\eea We will discuss the interpretation of $\Lambda_p;\Lambda_{max}$ in section (\ref{sec:discussion}).

\begin{figure}[H]
  \centering
  \includegraphics[width=\textwidth]{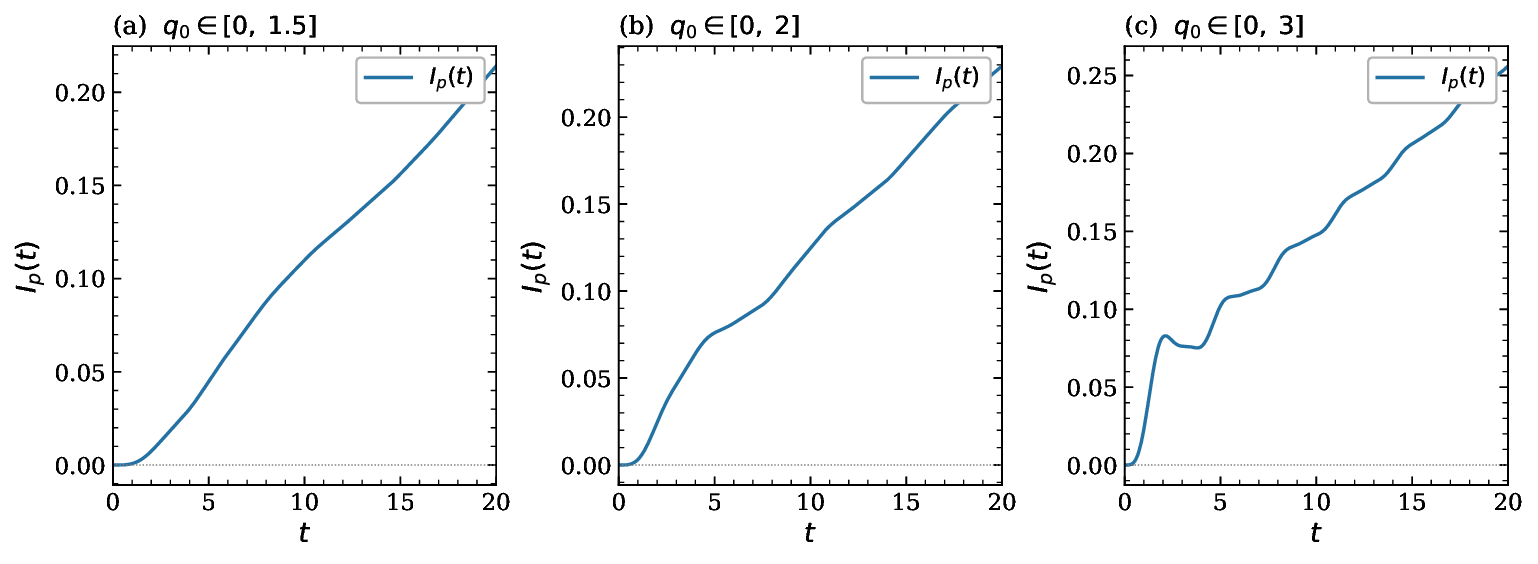}
  \caption{$I_p(t)$ for $\vq=0;\beta =1;t\in[0,20]$ with three upper cutoffs.
           (a)~$\Lambda_p=1.5$;\quad
           (b)~$\Lambda_p=2$;\quad
           (c)~$\Lambda_p=3$, all  in units where $m=1$.
           Dotted line: $I_p=0$. The early rise  is followed by oscillatory behavior that depends on the cutoff $\Lambda_p$ and is eventually followed by the linear time behavior from Fermi's Golden Rule. }
  \label{fig:Ipall}
\end{figure}

The region of integration of  $I_p(t)$  includes the resonant region at $q_0 \simeq E_{\phi}(\vq)$, this contribution features a rapid rise followed by oscillatory behavior that
depends on the upper cutoff $\Lambda_p$ which is followed by a linear secular growth in time in agreement with Fermi's Golden rule. This behavior is clearly shown in fig.(\ref{fig:Ipall}) which displays $I_p(t)$ vs $t$ for $\vq=0;\beta =1$ in units where $m=1$.

 In contrast, the contribution $I_c(t)$ does not contain the singular denominator as $q_0 \rightarrow E_\phi(\vq)$, however the zero temperature contribution to the spectral density (see equation (\ref{zeroTrho}) in appendix (\ref{app:ebcoup}))  behaves as $q^4_0$ for large $q_0$ therefore this integral  diverges strongly with the ultraviolet cutoff $\Lambda_{max}$.  In $I_c(t)$ we can safely separate the oscillatory cosine terms in $F[q_0,\vq,t]$ from the time independent non-oscillatory terms, namely
 \be I_c(t) = I_{osc}(t) + I_z \,,\label{oscisplit}\ee with
 \be I_z = 4\,E_\phi(\vq)\, \int^{\Lambda_{max}}_{\Lambda_p} \frac{q_0\, \rho(q_0,\vq)}{\big(q^2_0-E^2_\phi(\vq)\big)^2}\, \frac{dq_0}{2\pi}\,.\label{Iz}\ee

\begin{figure}[h]
  \centering
  \includegraphics[width=\textwidth]{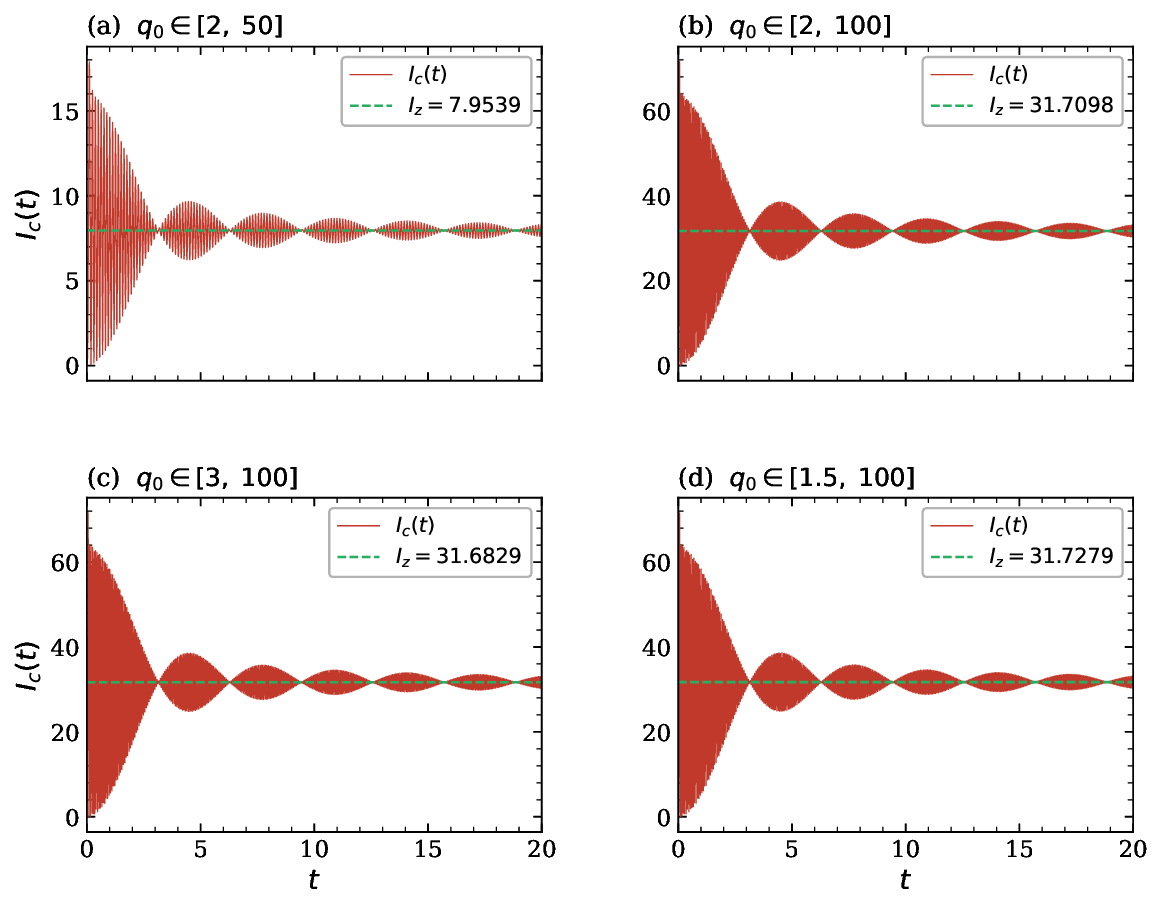}
  \caption{$I_c(t)$ (red) and $I_z$ (green dashed) for $\vq =0;\beta=1$ and  the four
           cutoff combinations:
           (a)~$[\Lambda_p,\Lambda_{max}]= [2,50]$;\quad(b)~$[2,100]$;\quad
           (c)~$[3,100]$;\quad(d)~$[1.5,100]$, in units where $m=1$.}
  \label{fig:Icall}
\end{figure}

Figure (\ref{fig:Icall}) displays $I_c(t)$ and $I_z$ vs. $t$ for $\vq =0;\beta=1$ in units where $m=1$, clearly showing that the early time behavior features strong oscillations associated with the ultraviolet cutoff $\Lambda_{max}$   averaging out on longer time scales to the asymptotic constant $I_z$.

 The large $q_0$ behavior of this integral is dominated by the zero temperature contribution to the spectral density (\ref{zeroTrho}) yielding a dependence on the ultraviolet cutoff $\Lambda_{max}$
 \be I_z \simeq E_{\phi}(\vq)\,\frac{\Lambda^2_{max}}{32\pi^2} \,, \label{largeIz}\ee explaining the factor $\simeq 4$ difference in the figures (\ref{fig:Icall}) with $\Lambda_{max} =50,100$.

 The asymptotic limit for large time and cutoff $\Lambda_{max}$ of the oscillatory component $I_{osc}(t)$ can be obtained straitghforwardly, it is dominated by the $q^4_0$ behavior of the zero temperature contribution to the spectral density, we find
 \be I_{osc}(t) = \frac{\Lambda^2_{max}}{32\pi^2 \,t}\,\Big[\cos\big(\Lambda_{max}\,t \big)\,\sin\big( E_{\phi}(\vq)\,t \big) - \frac{2 E_{\phi}(\vq)}{\Lambda_{max}} \,\sin\big(\Lambda_{max}\,t \big)\,\cos\big( E_{\phi}(\vq)\,t \big) \Big]+ \mathcal{O}(1/t^2) \,,\label{Iosi}  \ee which we have confirmed numerically and explains the slow envelope of the rapid oscillations and power law fall-off.

 Taken together, figures (\ref{fig:Ipall},\ref{fig:Icall})  confirm the asymptotic behavior (\ref{gamasy}), furthermore, they also confirm that the linear time dependence is actually \emph{subdominant} during a long period of time as the contribution from $I_c(t), I_z$ strongly depend on the cutoff and are much  larger than that of $I_p(t)$ during a long period of time.

 This analysis highlights the main features of the transient dynamics and identifies the contributions from the spectral density to the different behaviors:

 \begin{itemize}
  \item{The region $0\leq q_0 \leq \Lambda_p$ with $E_{\phi}(\vq) < \Lambda_p$ multiplying the function $F_+[q_0,\vq;t]$ gives rise to the long time secular term in agreement with Fermi's Golden rule along with a small constant in the long time limit. Within the same $q_0$ region, the contribution of $F_-[q_0,\vq;t]$ does not lead to a long time secular term, is purely oscillatory and remains small.}

  \item{ The region $E_{\phi}(\vq) < \Lambda_p \leq q_0 \leq \Lambda_{max}$ features two different behaviors: \textbf{a:)} a rapidly oscillatory contribution $I_{osc}(t)$ whose leading behavior is given by equation (\ref{Iosi}), averages out on time scales $t \simeq 2\pi/\Lambda_{max}$ and becomes negligible\footnote{Formally this is established by the averaging process on a time scale $\tau$ as $\frac{1}{\tau} \int^{\tau}_0 (\cos(\Lambda_{max} t') \cdots)dt'$ with $\Lambda_{max}\tau \gg 1$.} on time scales $\simeq 1/E_{\phi}(\vq)$, \textbf{b:)} an asymptotic constant term $I_z$ given by equation (\ref{largeIz}) which diverges $\propto \Lambda^2_{max}$ and yields the largest contribution to the quasiparticle residue $Z(\vq)$ (wave function renormalization).    }

      \item{ The contribution $I_c(t)$ (\ref{Icin})  that does \emph{not} feature the resonant denominator at $q_0 = E_\phi(\vq)$ in (\ref{Isplit}), and which in turn is split into the oscillatory component $I_{osc}(t)$ and the constant $I_z$ is such that $I_c(0) =0$, namely  $I_{osc}(0) = - I_z$. With $I_z$ \emph{defining} the quasiparticle weight  and $I_{osc}(t)$ averaging out on a short time scale $t_{qp}$  with $  2\pi/\Lambda_{max}\lesssim t_{qp} \ll  1/E_{\phi}(\vq)$ it follows that after this short time scale $\simeq t_{qp}$ the full contribution to $I_c(t)$ is determined by $I_z$ yielding the quasiparticle residue. The contribution $I_p(t)$ gives rise to the linear secular growth in time associated with the decay of the quasiparticle. We infer from this analysis that the short time scale $t_{qp}$ defines the \emph{formation} of the quasiparticle. In other words, the formation of the quasiparticle is a dynamical process  within this short time scale associated with the upper cutoff in the spectral density.  After its formation,  the quasiparticle decay is described by $I_p(t)$ which features the linear secular growth in time from Fermi's Golden rule at long time, but also includes  a non-secular purely oscillatory contribution from $F_-[q_0,\vq;t]$.    }

 \end{itemize}

 This analysis suggests that to separate more clearly the different behaviours it is instead  best to re-write the integral of the time dependent rate (\ref{gamasy})   as
 \be \int^t_0 \Gamma(\vq,t')\,dt' \equiv   {g^2} \, \mathcal{I}(t) \,,\label{intgam}\ee  where
 \be \mathcal{I}(t) \equiv \mathcal{I}_+(t)+ \mathcal{I}_{osc}(t)+ \mathcal{I}_Z \,,\label{maIsplit}\ee and the different contributions are given by
 \be \mathcal{I}_+(t) = \frac{1}{E_{\phi}(\vq)}\, \int^{\Lambda_p}_0 \rho(q_0,\vq)\,\Bigg[\frac{1-\cos\big[(q_0-E_{\phi}(\vq))t\big]}{( q_0-E_{\phi}(\vq))^2}\Bigg] \, \frac{dq_0}{2\pi}~~;~~ \Lambda_p > E_\phi(\vq)  \,,\label{Iplu}\ee including only $F_+[q_0,\vq;t]$,  leading to the linear secular term in time,
 \be \mathcal{I}_{osc}(t) = \frac{1}{E_{\phi}(\vq)}\,\int^{\Lambda_{max}}_0 \rho(q_0,\vq)\,\Bigg[\frac{\cos\big[(q_0+E_{\phi}(\vq))t\big]}{( q_0+E_{\phi}(\vq))^2}- \frac{\cos\big[(q_0-E_{\phi}(\vq))t\big]}{( q_0-E_{\phi}(\vq))^2}\,\Theta(q_0-\Lambda_p) \Bigg]\, \frac{dq_0}{2\pi} \,,\label{Iosci}\ee including all the non-secular and purely oscillatory contributions that average out on the short time scale $\simeq t_{qp}$, and
  \be \mathcal{I}_{Z} = \frac{1}{E_{\phi}(\vq)}\,\int^{\Lambda_{max}}_0 \rho(q_0,\vq)\,\Big[ \frac{\Theta(q_0-\Lambda_p)}{( q_0-E_{\phi}(\vq))^2}\, - \frac{1}{( q_0+E_{\phi}(\vq))^2}   \Big]\, \frac{dq_0}{2\pi} ~~; ~~ \mathcal{I}_{osc}(0) = - \mathcal{I}_Z \,,\label{Izit}\ee which determines the largest contribution to the quasiparticle ``residue''.

 The contribution from $\mathcal{I}_+(t)$ describes the \emph{effective low energy dynamics} of the quasiparticle, it contains the region $q_0 = E_\phi(\vq)$, it does \emph{not} depend on the high energy cutoff $\Lambda_{max}$  and  yields the linear secular growth in time at long time consistently with Fermi's Golden rule and exponential decay of the quasiparticle. It  features an early time rise $\propto t^2$ (from the early time behavior of the cosine term) in agreement with the result found in the seminal study in ref.\cite{fonda}. This early transient  is followed by an intermediate oscillatory  regime that depends crucially on the low energy cutoff $\Lambda_p$ and eventually   becomes the linear secular growth in time as per Fermi's Golden rule on time scales $ > 1/E_{\phi}(\vq)$.

  The early time transient $\propto t^2$
 is the precursor of  the quantum Zeno effect\cite{zeno1,zeno2,zeno3}: repeated projective measurements during this early transient regime results in inhibited decay as compared to the exponential decay of Fermi's Golden rule, an effect  which has been confirmed experimentally with ultra-cold atoms\cite{itano}. During this early transient
 \be g^2 \,\mathcal{I}_+(t)  \simeq \frac{t^2}{t^2_Z} ~~;~~ \frac{1}{t^2_Z} = \frac{g^2 }{4E_\phi(\vq)}\,\int^{\Lambda_p}_0 \rho(q_0,\vq) \, \frac{dq_0}{2\pi}\,,\label{zenog}\ee where the Zeno time scale $t_Z$ depends strongly on $\Lambda_p$ as a consequence of the rapid growth of the spectral density.

 However, when the $\propto t^2$ growth becomes larger than the linear growth from Fermi's Golden rule, the decay is instead \emph{accelerated}, this is the quantum anti-Zeno effect\cite{zeno3} which results in enhanced, rather than suppressed decay. The criterion for anti-Zeno enhancement is that $t> t_c$ with $t_c$ the crossover time from the $\propto t^2/t^2_Z$ behavior to $\Gamma(\vq;\infty) t$ (Fermi's Golden rule) behavior, namely
 \be t > t_c = \Gamma(\vq;\infty) \,  t^2_Z  \,, \label{crosstim}\ee with the constraint that $t_c$ must still be small enough that the $t^2$ behavior describes the transient dynamics. Whether this behavior emerges   depends on the details of the spectral density and $\Lambda_p$ (note that $t_c$ is independent of $g^2$).

\begin{figure}[htbp]
  \centering
  \includegraphics[width=0.45\textwidth]{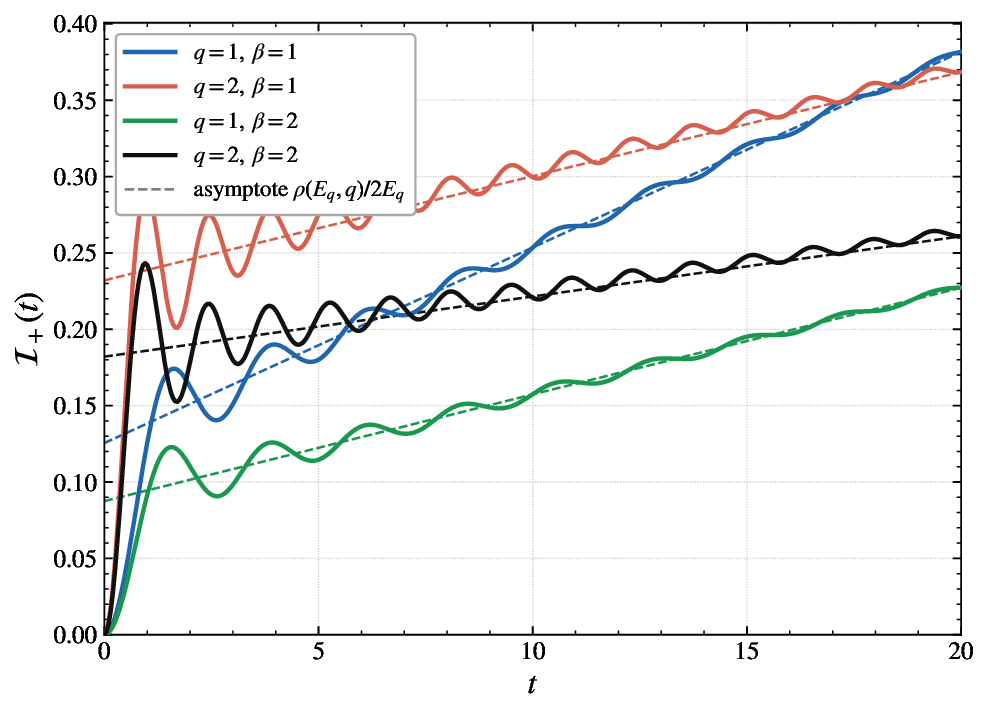}
  \hfill
  \includegraphics[width=0.45\textwidth]{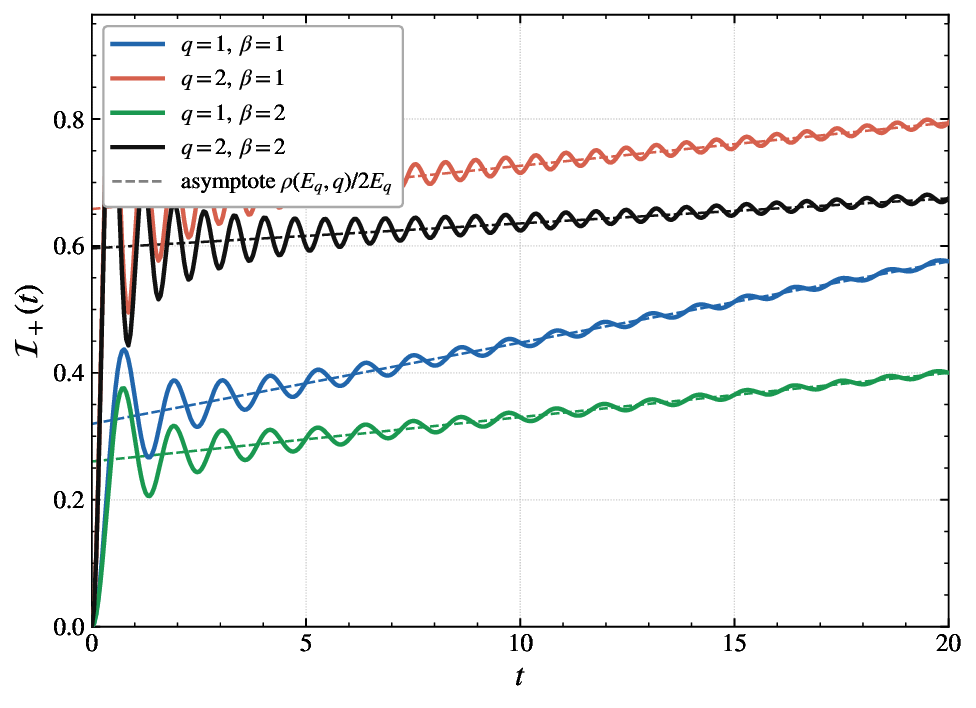}
   \caption{$\mathcal{I}_+(t)$ with $E_{\phi}(\vq) \equiv \sqrt{q^2+1}$, for $q=1,2;\beta=1,2$ in units where $m=1$ for $\Lambda_p = 3 E_\phi(q)$ (left panel) and $\Lambda_p = 5 E_\phi(q)$ (right panel). The dashed lines are asymptotes  $z_0+ \frac{\rho(E_{\phi}(q))}{2E_\phi(q)}\,t$, their intercepts $z_0$ contribute to the quasiparticle residue. The intersects and oscillation frequencies   depend on $\Lambda_p$. Early time transient is characterized by quantum anti-Zeno enhancement. }
  \label{fig:iplus}
\end{figure}

   Figures (\ref{fig:iplus}) show this behavior with the full spectral density (\ref{rhofi}) for values $q=1,2;\beta = 1,2;\Lambda_p = 3 E_{\phi}(\vq),  5 E_{\phi}(\vq)$ for a dispersion relation $E_{\phi}(\vq)= \sqrt{q^2+1}$ all in units where $m=1$. The dashed lines in this figure display a linear fit $z_0 + \frac{\rho(E_{\phi}(q))}{2E_\phi(q)}\,t$.   The intercept at $t=0$, namely $z_0$, is a small contribution to  $Z(q)$ the quasiparticle residue that depends on $\Lambda_p$ but not on $\Lambda_{max}$, and the crossover  time scale $t_c$ from Zeno to anti-Zeno behavior is shorter for larger $\Lambda_p$, as evidenced by these figures,  because of the growth of the spectral density $\simeq q^4_0$, a distinct feature of the Chern-Simons coupling.

These figures are representative of an exhaustive numerical study which confirms that for $\Lambda_p > E_{\phi}(\vq)$ the quantum Zeno suppression is restricted to a very short
time scale, but \emph{anti}-Zeno enhancement of the decay is actually the feature that dominates the transient dynamics: the effective time dependent decay function $\mathcal{I}_+(t)$  is always above the straight line with slope $\frac{\rho(E_{\phi}(q))}{2E_\phi(q)}$ and zero intercept   corresponding to Fermi's Golden rule, as can be understood from the fact that the intercept $z_0>0$ in all cases.

This is an important result of our study: early time transient dynamics features quantum anti-Zeno\cite{zeno3} behavior with enhanced decay as compared to Fermi's Golden rule.

The oscillatory part $\mathcal{I}_{osc}(t)$ features the asymptotic behavior (\ref{Iosi}) (divided by $E_{\phi}(\vq)$) with strong oscillations with a time scale $\simeq 2\pi/\Lambda_{max}$, with an envelope that evolves on a much slower time scale $\propto 1/E_\phi(\vq)$ and a power fall off $\propto 1/t$.  This contribution    averages out  rapidly by dephasing (Riemann-Lebesgue)   and is similar to that displayed in fig. (\ref{fig:Icall}).

  Finally,  $\mathcal{I}_Z = \frac{\Lambda^2_{max}}{32\pi^2}  + \cdots$ where the dots stand for subleading terms in the large $\Lambda_{max}$ limit, is the long time asymptote from the contributions from the spectral density from regions of $q_0$ far away from the resonant denominator and is the largest contribution to the quasiparticle ``residue'' $Z(\vq)$.   $g^2 \mathcal{I}_Z$ differs from the quasiparticle residue $Z(\vq)$   by $g^2 z_0$, where  $z_0$ are the intercepts from $\mathcal{I}_+(t)$, these terms   are perturbatively small ($\propto g^2$)   and, crucially,  independent of the high energy cutoff $\Lambda_{max}$.

This analysis suggests to define
\be \gamma(t) = g^2 \Big[\mathcal{I}_+(t)+\mathcal{I}_{osc}(t) \Big] \,,\label{lilgam}\ee it follows that the decay function that enters in the solutions (\ref{aoftsol},\ref{aaoftsol}) can be written as
\be e^{-\int^t_0 \Gamma(\vq,t')\,dt'} = \mathcal{Z}(\vq)\,e^{-\gamma(t)} \ee where
\be \gamma(t)~~{}_{\overrightarrow{t\gg t_{qp}}}~~ \, g^2\, \mathcal{I}_+(t) \,,\label{asytqp}\ee and
\be \mathcal{Z}(\vq) = e^{-g^2 \mathcal{I}_Z}\simeq 1- g^2 \,\mathcal{I}_Z +\cdots \,. \label{capZ}\ee  differs from the quasiparticle residue $Z(\vq)$ (\ref{wfun}) by  $g^2 z_0$. This expression makes explicit that the contribution $\mathcal{I}_+(t)$ effectively describes the decay of the quasiparticle on time scales longer than its ``formation time'' $t_{qp}$.

\subsection{Population dynamics:}\label{sub:popdyn}

The population $N(\vq,t)$ given by eqn. (\ref{disfun}) becomes
\be N(\vq,t) = \mathcal{Z}(\vq)\,N(\vq,0)\,e^{-\gamma(t)} + \mathcal{N}(\vq,t) \,,\label{pop1}\ee where
 using the property $\rho(q_0;\vq) = -\rho(-q_0;\vq)$, writing
\be \Gamma^< (\vq,t) = \frac{g^2}{E_{\phi}(\vq)} \, \int^{\Lambda_{max}}_0 \rho(q_0;\vq)\,\Big[n(q_0)\,  \frac{\sin[(E_\phi(\vq)-q_0)\,t]}{(E_\phi(\vq)-q_0)}+ \big(1+n(q_0) \big)\frac{\sin[(E_\phi(\vq)+q_0)\,t]}{(E_\phi(\vq)+q_0)}\Big]\,\frac{dq_0}{2\pi}  \,,\label{Gamgi}\ee and separating the region that contains the resonance at $q_0 = E_\phi(\vq)$ as in the analysis above, it follows that

\be \mathcal{N}(\vq,t) = \mathcal{N}_p(\vq,t) + \mathcal{N}_{osc}(\vq,t) \,,\label{Nsplit}\ee where
\be \mathcal{N}_p(\vq,t) =  \frac{g^2}{E_{\phi}(\vq)} \, \int^t_0 \Bigg\{  \int^{\Lambda_{p}}_0 \rho(q_0;\vq)\,n(q_0)\,  \frac{\sin[(E_\phi(\vq)-q_0)\,t']}{(E_\phi(\vq)-q_0)}\,\frac{dq_0}{2\pi}\Bigg\}\,e^{-\big(\gamma(t)-\gamma(t')\big)}\,dt'  \,,\label{Nplu}\ee and
\bea \mathcal{N}_{osc}(\vq,t) &  = &  \frac{g^2}{E_{\phi}(\vq)} \, \int^t_0  \Bigg\{ \int^{\Lambda_{max}}_0 \rho(q_0;\vq) \Bigg[ n(q_0)\,  \frac{\sin[(E_\phi(\vq)-q_0)\,t']}{(E_\phi(\vq)-q_0)}\,\Theta(q_0-\Lambda_p) \nonumber \\ & + & \big(1+n(q_0) \big)\frac{\sin[(E_\phi(\vq)+q_0)\,t']}{(E_\phi(\vq)+q_0)}\Bigg]\,\frac{dq_0}{2\pi} \Bigg\}\,e^{-\big(\gamma(t)-\gamma(t')\big)} \,dt'
\,.\label{Nos}\eea We highlight that whereas the first term in equation (\ref{pop1}) depends on the quasiparticle weight, the second \emph{does not}, therefore, because  the first term
vanishes asymptotically ($\gamma(t) \rightarrow \Gamma(\vq,\infty)\,t$), the final population at long time $t \gg 1/\Gamma(\vq,\infty)$  does not depend on the quasiparticle residue.

\begin{figure}[htbp]
  \centering
  \includegraphics[width=0.45\textwidth]{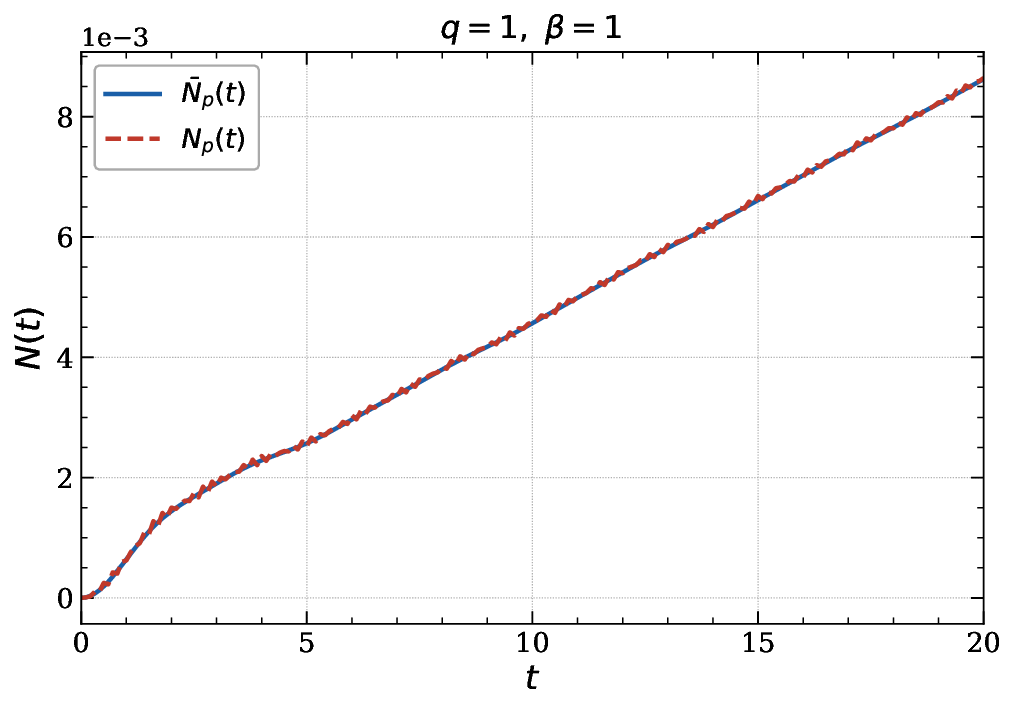}
  \hfill
  \includegraphics[width=0.45\textwidth]{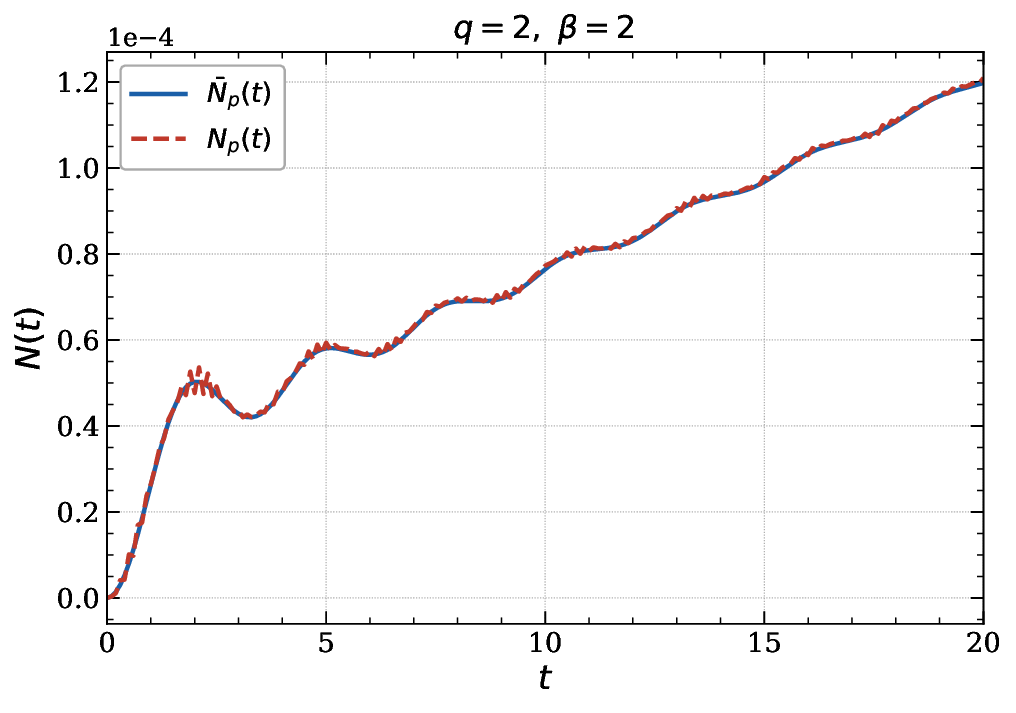}
   \caption{$\mathcal{N}_p(t)$ with $E_{\phi}(\vq) \equiv \sqrt{q^2+1}$, for $q=1,2;\beta=1,2$ in units where $m=1$ for $\Lambda_p = 3 E_\phi(q);\Lambda_{max}=50$, $g^2 =0.1$  with and without ($\overline{\mathcal{N}}_p(t)$)  the oscillatory contribution to the damping factor $\gamma(t)$. The curves are indistinguishable after a time scale $t \simeq 1/E_{\phi}(\vq)$. }
  \label{fig:pop1}
\end{figure}

Before we proceed to a numerical study, we can gain insight into the behavior of  both components by using the results of the analysis of $\mathcal{I}_+(t)$ above, and replace
\be \gamma(t) \simeq g^2\,z_0 + \Gamma(\vq,\infty) t \,,\label{gamsim}\ee thereby neglecting the small oscillatory behavior displayed in fig.(\ref{fig:iplus}), allowing us to carry out the time integrations explicitly. For the previous numerical examples, with $E_\phi(\vq) = \sqrt{q^2+1}$ the time scale for decay is approximately given by
\bea  && q = 1;\beta =1, g^2 = 0.1 \Rightarrow \Gamma(\vq;\infty) \simeq 1.28 \times 10^{-3} ~~;~~ \frac{1}{\Gamma(\vq;\infty)} \simeq 780 \nonumber \\
&& q = 2;\beta =2, g^2 = 0.1 \Rightarrow \Gamma(\vq;\infty) \simeq 3.94 \times 10^{-4} ~~;~~ \frac{1}{\Gamma(\vq;\infty)} \simeq 2537 \,,\label{values}\eea all in units there $m=1$. Therefore on time scales $t \ll 1/\Gamma(\vq;\infty)$ we can effectively set the damping exponentials in (\ref{Nplu},\ref{Nos}) $e^{-(\gamma(t) - \gamma(t'))}\simeq 1$, and the resulting integrals are very similar to $\mathcal{I}_+(t)$ (\ref{Iplu}) and $\mathcal{I}_{osc}(t)+\mathcal{I}_Z$, (\ref{Iosci}, \ref{Izit}) studied previously (but multiplied by $g^2$) with the only difference being the thermal factors but with qualitatively the same temporal behavior. Therefore $\mathcal{N}_p(t) $  is expected to raise rapidly $\propto t^2$ at very early time and grow linearly up to about a time scale $1/\Gamma(\vq;\infty)$ at which the exponential damping will become important. This expectation is supported by the numerical analysis shown in fig. (\ref{fig:pop1}) which does include the exponential damping in (\ref{Nplu}) during this early stage. The oscillatory contribution $\mathcal{I}_{osc}(t)$ to $\gamma(t)$ in (\ref{lilgam}) averages out on a short time scale of quasiparticle formation $2\pi/\Lambda_{max} \lesssim t_{qp} \ll 1/E_{\phi}(\vq)$  and can be neglected in numerical studies of the population, this is explicitly shown in fig. (\ref{fig:pop1}) that displays $\mathcal{N}_p(t)$ including its contribution and $\overline{\mathcal{N}}(t)$ without it for various values of $q,\beta$.

In the asymptotic long time limit
\be \mathcal{N}_p(\vq,t)_{~~\overrightarrow{t\rightarrow \infty}}~\,n\big(E_{\phi}(\vq)\big)\,\big(1-e^{-\Gamma(\vq;\infty)\,t}\big) \,,\label{Npluas}\ee  because in this limit
\be \gamma(t) \rightarrow \Gamma(\vq;\infty) t~;~ \Gamma^<(\vq;t) \rightarrow \Gamma(\vq;\infty)\,n(E_\phi(\vq)) \,,\label{asymod}\ee
 however at early time the population grows $\propto t^2$ which  can be understood from the fact that at early time the damping factor in (\ref{Nplu}) is negligible, therefore
\be  \mathcal{N}_p(\vq,t) \simeq   \frac{g^2}{E_{\phi}(\vq)} \,      \int^{\Lambda_{p}}_0 \rho(q_0;\vq)\,n(q_0)\,  \Big[\frac{ 1-\cos\big[(q_0-E_{\phi}(\vq))t\big]}{( q_0-E_{\phi}(\vq))^2} \Big] \,\frac{dq_0}{2\pi} \,.\label{earlynp}     \ee

\begin{figure}[htbp]
  \centering
  \includegraphics[width=0.45\textwidth]{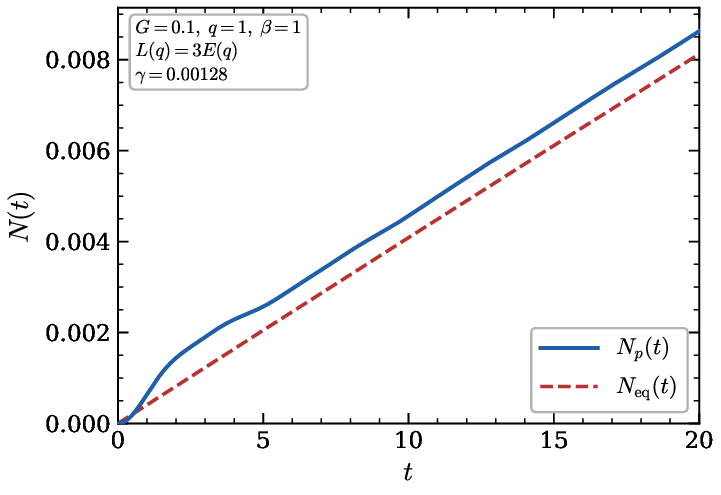}
  \hfill
  \includegraphics[width=0.45\textwidth]{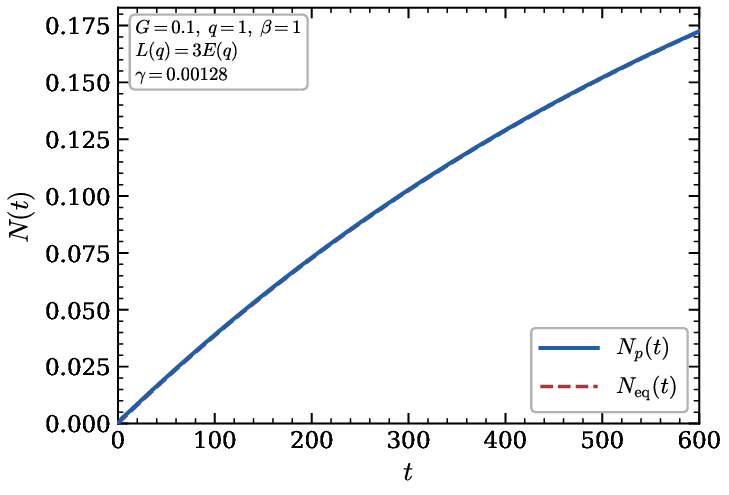}
   \caption{$\mathcal{N}_{p}(t)$ with $E_{\phi}(\vq) \equiv \sqrt{q^2+1}$, for $q=1 ;\beta=1 $ in units where $m=1$ for $\Lambda_p = 3 E_\phi(q);\Lambda_{max}=50;G= g^2 =0.1$ and $n\big(E_{\phi}(\vq)\big)\,\big(1-e^{-\Gamma(\vq;\infty)\,t}\big)$ with $\Gamma(\vq;\infty) = 0.00128$. At early time the population grows faster than with Fermi's Golden rule and detailed balance: quantum anti-Zeno effect. }
  \label{fig:npneq}
\end{figure}

 Early and late time dynamics are displayed in fig. (\ref{fig:npneq}) which shows the asymptotic limit (\ref{Npluas}) and departure at early time when the population grows \emph{faster} than that predicted by Fermi's Golden rule and detailed balance $\simeq n\big(E_{\phi}(\vq)\big)\, \Gamma(\vq;\infty)\,t$ approaching the  asymptotic behavior (\ref{Npluas}) at long time.

The origin of the early time transient $\propto t^2$ and its rise above the straight line   from Fermi's Golden rule and detailed balance as shown in the left panel of fig. (\ref{fig:npneq}) is similar to that of $\mathcal{I}_+(t)$ (\ref{maIsplit}),  it is a manifestation  of the quantum anti-Zeno effect\cite{zeno1,zeno2,zeno3,itano}, imprinted   in the population dynamics as a faster build up of the population. However the effective Zeno time scale for the build-up of the population is now weighted by the thermal distribution function.

\begin{figure}[htbp]
  \centering
  \includegraphics[width=0.45\textwidth]{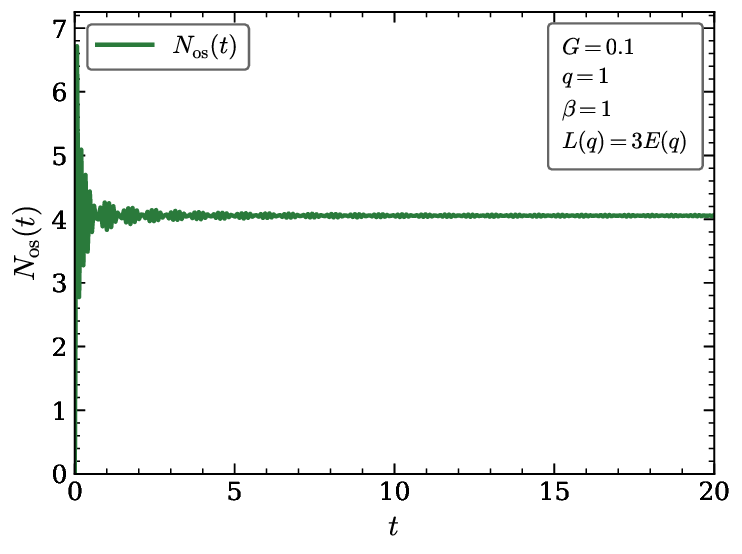}
  \hfill
  \includegraphics[width=0.45\textwidth]{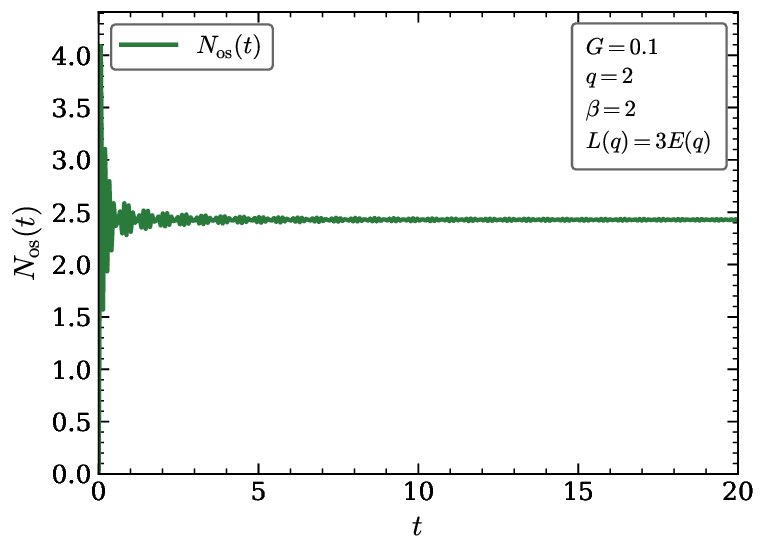}
   \caption{$\mathcal{N}_{osc}(t)$ with $E_{\phi}(\vq) \equiv \sqrt{q^2+1}$, for $q=1,2;\beta=1,2$ in units where $m=1$ for $\Lambda_p = 3 E_\phi(q);\Lambda_{max}=50$   ; $G= g^2 =0.1$ .}
  \label{fig:nosc}
\end{figure}

A similar analysis applies to the oscillatory component (\ref{Nos}). On a time scale $t \ll 1/\Gamma(\vq,\infty)$ we can effectively neglect the damping factor and the time integral can be carried out, yielding an expression similar to the sum $\mathcal{I}_{osc}(t)+ \mathcal{I}_Z$ (\ref{Iosci}, \ref{Izit}), (up to the coupling $g^2$) differing solely in the thermal distributions. Therefore on this time scale we expect a similar behavior, a rapid rise with averaging on a time scale of the order of $t_{qp}$ with a nearly constant asymptote, the counterpart of $\mathcal{I}_Z$. This expectation is borne out by the numerical study including the damping factor, displayed in fig. (\ref{fig:nosc}).

\begin{figure}[htbp]
  \centering
  \includegraphics[width=0.45\textwidth]{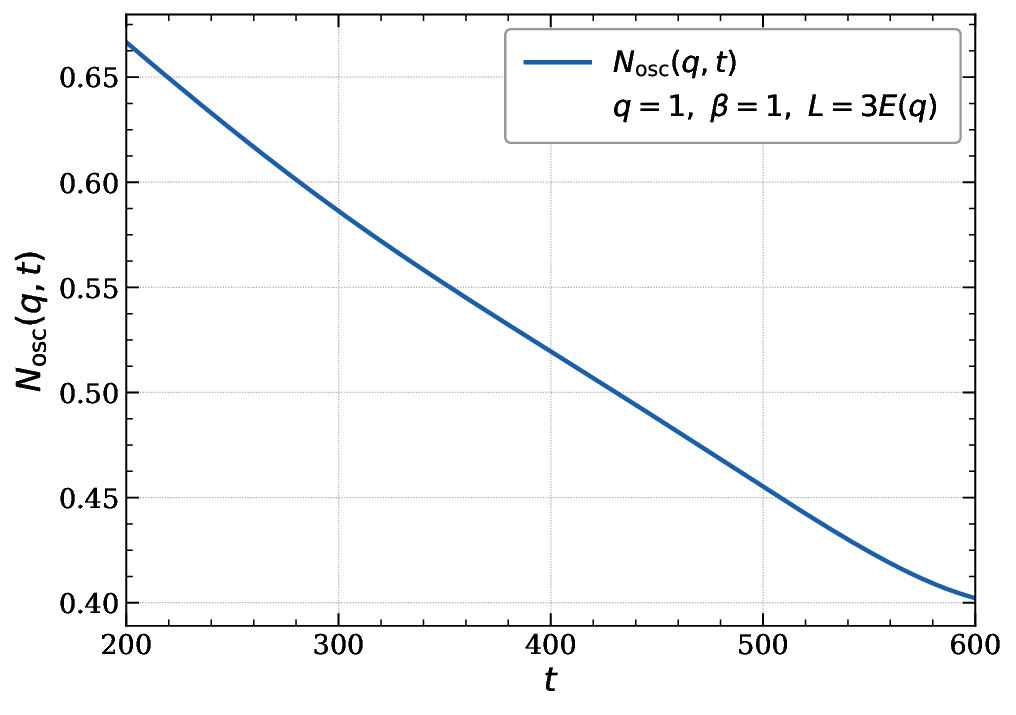}
  \hfill
  \includegraphics[width=0.45\textwidth]{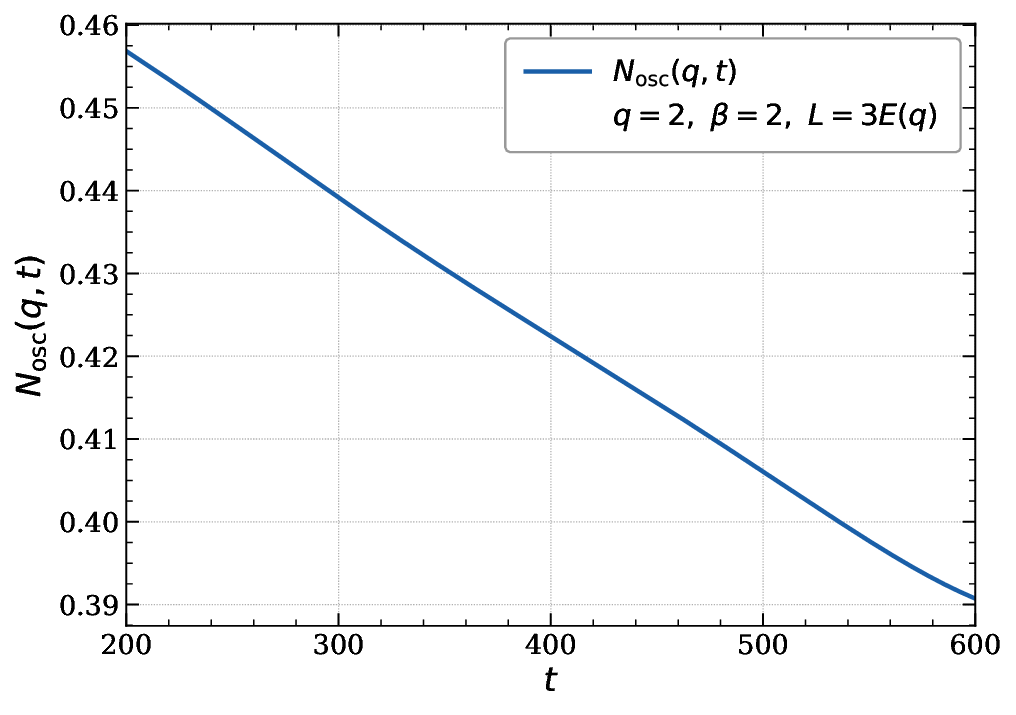}
   \caption{$\mathcal{N}_{osc}(t)$ with $E_{\phi}(\vq) \equiv \sqrt{q^2+1}$, for $q=1,2;\beta=1,2$ in units where $m=1$ for $\Lambda_p = 3 E_\phi(q);\Lambda_{max}=50$   ; $G= g^2 =0.1$ .}
  \label{fig:nosasy}
\end{figure}

On longer time scales, we can implement the simplification (\ref{gamsim}) and the time integrals can be done straightforwardly, with the result
\be    \mathcal{N}_{osc}(\vq,t)~~ {}_{\overrightarrow{t \gg t_{qp}}}~~   \, e^{-\Gamma(\vq,\infty)t}\,   \frac{g^2}{E_{\phi}(\vq)} \,  \int^{\Lambda_{max}}_0 \rho(q_0;\vq) \Bigg[    \frac{n(q_0)\,\Theta(q_0-\Lambda_p)}{(E_\phi(\vq)-q_0)^2+\Gamma^2(\vq,\infty) }   +  \frac{ \big(1+n(q_0) \big)}{(E_\phi(\vq)+q_0)^2+\Gamma^2(\vq,\infty)}\Bigg] \,\frac{dq_0}{2\pi}  \,. \label{nosasys}  \ee where the rapid oscillations have averaged out, this asymptotic limit is confirmed numerically and displayed in fig.(\ref{fig:nosasy}).

In summary: population kinetics differs from the usual approach to equilibrium from Fermi's Golden rule and detailed balance in early time transient behavior with faster population build-up as a consequence of the quantum anti-Zeno effect and violation of detailed balance. The early time transient  is followed at intermediate and late time by an   asymptotic approach to equilibrium that merges with the Fermi's Golden rule result (\ref{Npluas}) but includes the perturbatively small   contribution (\ref{nosasys}) that decays on the time scale $1/\Gamma(\vq;\infty)$, which is a remnant of the quasiparticle formation.

\vspace{1mm}

\subsection{Early time violation of detailed balance:}\label{sub:detbal}

\begin{figure}[htbp]
  \centering
  \includegraphics[width=0.45\textwidth]{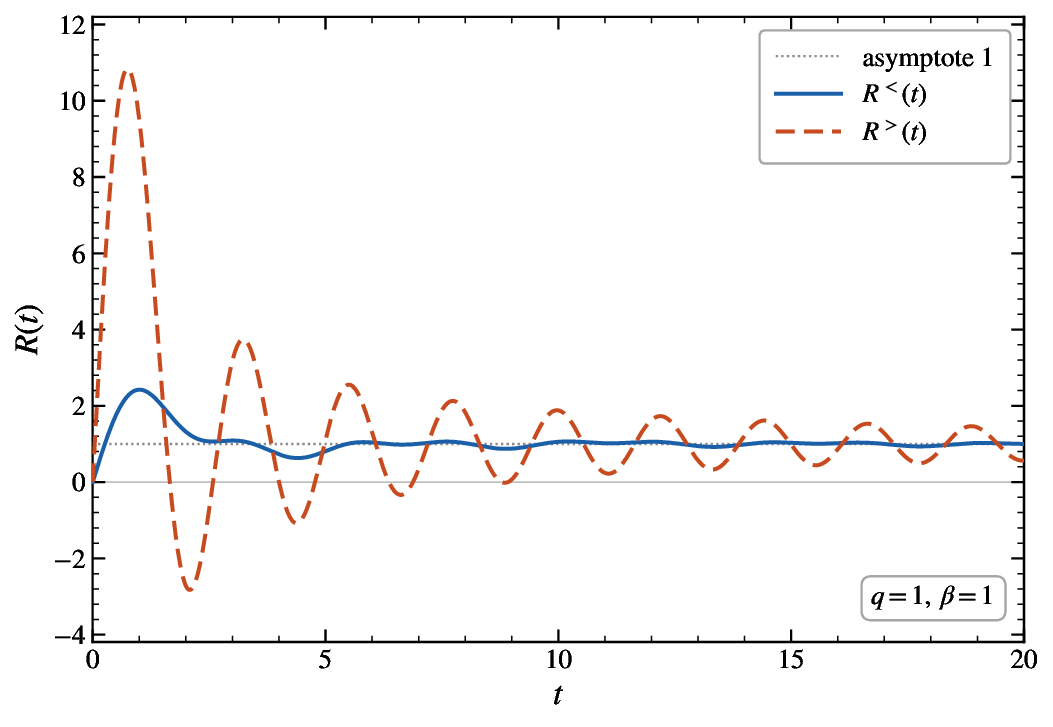}
    \caption{Ratios $R^<(t)= \frac{\Gamma^>(\vq,t)}{\Gamma(\vq,\infty) n(E_\phi(\vq)}$, and  $R^>(t)= \frac{\Gamma^>(\vq,t)}{\Gamma(\vq,\infty)(1+n(E_\phi(\vq))}$ for $q=1 ;\beta=1 $ in units where $m=1$ for $\Lambda_p = 3 E_\phi(q)$ neglecting the oscillatory contribution  $\mathcal{N}_{osc}(t)$.}
  \label{fig:detbal}
\end{figure}

The time dependent forward (decay) and backward (recombination) rates obey the detailed balance relation (\ref{detbal}) in the asymptotic long time limit, however this relation is \emph{not} fulfilled at early time. Both rates grow linearly in time and begin to oscillate and it is only in the long time limit that the formal identity (\ref{sinlt}) holds  and detailed balance is fulfilled. The linear rise in time of both rates is responsible for the early time $t^2$ behavior of both the decay function $\int^t_0 \Gamma(\vq,t') dt'$  and the population $\mathcal{N}_p(t)$. As discussed above this early time behavior is a hallmark of the quantum zeno effect\cite{fonda,zeno1,zeno2}, which has been experimentally confirmed\cite{itano}  but as consequence of the particular spectral density is manifest as an enhancement of decay or population build-up, namely the quantum anti-Zeno effect\cite{zeno3} . Therefore we conclude that the early time violation of detailed balance is a manifestation of the quantum anti-zeno effect. Fig.(\ref{fig:detbal}) displays the ratios $R^>(t) = \Gamma^>(\vq,t)/(\Gamma(\vq,\infty) n(E(q));R^<(t) = \Gamma^>(\vq,t)/(\Gamma(\vq,\infty) (1+ n(E(q)))$ neglecting the rapidly averaging oscillatory component  $\mathcal{N}_{osc}(t)$. This figure shows clearly the early linear rise in time followed by oscillatory behavior that at long time merges with the detailed balance asymptotic limits. In particular $R^>(t)$ features much larger oscillations as a consequence of the ``spontaneous'' contribution (vacuum term) which is not thermally suppressed and allows to probe a larger region of the spectral density during the early time transient.

The conclusion of this analysis is that the early time transients which are the hallmark of the quantum zeno effect, translate into early time violations of detailed balance.

Early and late time transients along with the contribution from $\mathcal{I}_Z$ which determines the quasiparticle residue are all corrections to the usual Fermi's Golden rule. Departures from and violations of Fermi's Golden Rule including oscillatory dynamics superimposed with exponential decay have been    observed experimentally in various systems\cite{expdecay,navon,wilk,raizen}, and the dynamics of formation of a quasiparticle from an impurity in a Fermi sea has been studied with ultrafast interferometry\cite{cetina,viva}. Perhaps similar experimental techniques with ultra-fast spectroscopy and femtosecond time resolutions may be harnessed to probe the transient dynamics of  (DAQ) excitations and   reveal quantum anti-Zeno dynamics in topological materials.

 \section{Condensate equation of motion.}\label{sec:eomlr}
\begin{figure}[h]
  \centering
  \includegraphics[width=3in]{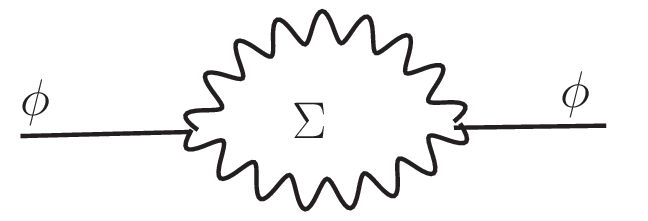}
  \caption{$\phi$ one photon loop self energy.}
  \label{fig:fise}
\end{figure}

 We now obtain the equation of motion for the expectation value of the (DAQ) field (condensate) by using the time evolution of the reduced   density matrix  in interaction picture (\ref{Linblad}) \emph{before} invoking the Markov approximation $\rho_{I\phi}(t') \rightarrow \rho_{I\phi}(t)$ and the rotating wave approximation.  The Hamiltonian equations of motion for the Heisenberg operator (in the Heisenberg picture the density matrix does not depend on time)
 \be \dot \phi(\vec{y},t) = \pi(\vec{y},t) \Rightarrow \frac{\partial}{\partial t} \langle \phi(\vec{y},t) \rangle = \mathrm{Tr}\, \pi_I(t) \rho_I(t) \,,  \label{heiseom}\ee where we used the identity (\ref{Oex}).
 Taking another time derivative
 \be \frac{\partial^2}{\partial t^2} \langle \phi(\vec{y},t) \rangle = \mathrm{Tr}\, \big[ \dot{\pi}_I(t) \rho_I(t) + \pi_I(t) \dot{\rho}_I(t)\big] \,,\label{eom2}\ee
 with the interaction picture (free field) equation of motion
 \be \dot{\pi}_I(\vy,t) = \big(\vec{v}\cdot\vec{\nabla}\big)\big(\vec{v}\cdot\vec{\nabla}\phi_I(\vy,t)\big)  - m^2 \phi_I(\vy,t) \,,\label{ffeom}\ee it remains to obtain the second term in the bracket in (\ref{eom2}), and since $\pi_I$ only depends on (DAQ) degrees of freedom, the trace over the bath variables can be done yielding the reduced density matrix whose equation of motion is given by (\ref{Linblad}).   Using the cyclic property of the trace, along with the canonical commutation relation $[\pi_I(\vy,t),\phi_I(\vx,t)] = -i\,\delta^{(3)}(\vy-\vx)$, and the identity (\ref{Oex})  we obtain
 \be \ddot{\overline{\phi}}(\vy,t)-\big(\vec{v}\cdot\vec{\nabla}\big)\big(\vec{v}\cdot\vec{\nabla}\overline{\phi}(\vy,t)\big)+m^2\,\overline{\phi}(\vy,t) + \int^t_0 dt' \int d^3 x\,\Sigma(\vy-\vx,t-t')\,\overline{\phi}(\vx,t') = 0 ~~;~~  \overline{\phi}(\vy,t) \equiv  \langle \phi(\vec{y},t) \rangle  \,,\label{reteom}  \ee where
 \be \Sigma(\vy-\vx,t-t')= -i g^2 \Big[ G^>(\vy-\vx,t-t') - G^<(\vy-\vx,t-t')\Big] \,,\label{selfen}\ee is the retarded proper (one particle irreducible) one loop self-energy\cite{fetter,mahan} from two-photon exchange \footnote{The $\Theta(t-t')$ in the retarded proper self-energy is accounted for by the upper limit in the time integral in (\ref{reteom})} corresponding to the Feynman diagram depicted in fig.(\ref{fig:fise}). Using the relations (\ref{ggreat},\ref{gless}) it follows that the self energy can be written as the commutator
 \be \Sigma(\vy-\vx,t-t') = -i g^2 \, \underset{\{\gamma\}}{\text{Tr}} \,\rho_\gamma \, \Big[ \mathcal{C}_I(\vy,t) \,, \mathcal{C}_I(\vx,t')\Big]  \,.\label{commu}\ee Using the spectral representations (\ref{Ggfd},\ref{Glfd}) and taking the spatial Fourier transform of $\overline{\phi}(\vy,t)$ the equation of motion (\ref{reteom}) becomes
 \be \ddot{\Phi}(\vq,t)+ E^2_{\phi}(\vq)\,\Phi(\vq,t) + \int^t_0 \sigma(\vq;t-t') \,\Phi(\vq,t')\,dt' =0 \,,\label{FTeom}\ee where $\Phi(\vq,t)$ is the spatial Fourier transform of $\overline{\phi}(\vy,t)$ and
 \be \sigma(\vq;t-t') = -i\,g^2\, \int \rho(q_0,\vq) \,e^{-iq_0(t-t')}\,\frac{dq_0}{2\pi} \,,\label{SigmaFT} \ee is the spatial Fourier transform of the self-energy,  where the spectral density $\rho(q_0,\vq)$ is given by eqn. (\ref{specOs}). The equation of motion (\ref{FTeom}) can be solved by Laplace transform:  introducing the Laplace transforms
\bea \widetilde{\Phi}(\vq;s) & = &  \int^\infty_0 e^{-st}\,\Phi(\vq,t)\,dt \,,\label{laplaX}    \\ \widetilde{\sigma}(s;\vq) & = &  \int^\infty_0 e^{-st}\,\sigma(\vq,t)\,dt  = -  g^2 \,\int^{\infty}_{-\infty} \frac{\rho(q_0,\vq)}{q_0-is} \frac{dq_0}{2\pi} \,.\label{laplasigma}
\eea

the solution of the Laplace transform   is
 \be \widetilde{\Phi}(\vq;s) = \frac{\dot{\Phi}(\vq,0) +s\,\Phi(\vq,0)}{s^2+E^2_{\phi}(\vq)+\widetilde{\sigma}(s;\vq)}\,.\label{laplasolution}\ee
The solution in real time is obtained by inverse Laplace transform, it is given by
\be \Phi(\vq,t)   =  \Phi(\vq,0)\,\dot{\mathcal{G}}(\vq;t) + \dot{\Phi}(\vq,0)\,\mathcal{G}(\vq;t)   \,,\label{realtisol} \ee  where $\mathcal{G}(\vq,t)$ is given by
\be \mathcal{G}(\vq;t) = \frac{1}{2\pi i} \int_{ \mathbb{{C}}} \frac{e^{st}}{s^2+E^2_\phi(\vq)+\widetilde{\sigma}(s;\vq)}\, ds \,, \label{goftsol}\ee   $ \mathbb{{C}}$ denotes the Bromwich contour parallel to the imaginary axis   and to the right of all the singularities   of $(s^2+E^2_\phi(\vq)+\widetilde{\sigma}(\vq;s))^{-1}$ in the complex s-plane, and closing along a large semicircle at infinity with $Re(s)<0$. These singularities correspond to poles and multiparticle branch cuts with $Re(s)<0$, thus the contour runs parallel to the imaginary axis $s= i(\omega -i \epsilon)$, with $-\infty \leq \omega \leq \infty$ and $\epsilon \rightarrow 0^+$. Therefore,
\be \mathcal{G}(\vq;t) = - \int^{\infty}_{-\infty} \widetilde{\mathcal{G}}(\omega;\vq)\, {e^{i\omega\,t}} \,\frac{d\omega}{2\pi}\,, \label{Goftfin}\ee  where
\be \widetilde{\mathcal{G}}(\omega;\vq) = \frac{1}{(\omega-i\epsilon)^2 -E^2_{\phi}(\vq) - \widetilde{\sigma}(\omega;\vq) }\,, \label{Gfnu}  \ee is recognized as the retarded Dyson propagator, namely the geometric series of the proper self-energy insertions\cite{fetter,mahan}.
The self energy in frequency space is given by the dispersive form
\be \widetilde{\sigma} (\omega;\vq)   =    g^2\,\int^{\infty}_{-\infty}  \,  \frac{\rho(q_0,\vq)}{\omega-q_0-i\epsilon} \,\frac{dq_0}{2\pi} \equiv \widetilde{\sigma}_R(\omega;\vq) + i \, \widetilde{\sigma}_I( \omega;\vq) \,,
 \label{signu}\ee with the real and imaginary parts given by
 \bea \widetilde{\sigma}_R(\omega;\vq) & = &   g^2\, \int^{\infty}_{-\infty} \mathcal{P}\,  \Bigg[ \frac{\rho(q_0,\vq)}{\omega-q_0}\Bigg]\,\frac{dq_0}{2\pi} \,,\label{resig}\\
  \widetilde{\sigma}_I(\omega;\vq) & = & \frac{g^2}{2}\,\rho(\omega,\vq) \,, \label{imsig} \eea satisfying  the Kramers-Kronig relation
  \be \widetilde{\sigma}_R(\omega;\vq) = \frac{1}{\pi} \,\int^{\infty}_{-\infty} \mathcal{P}\,  \Bigg[ \frac{ \widetilde{\sigma}_I(q_0;\vq)}{\omega-q_0}\Bigg]\,  {dq_0} \,.\label{KKrela}\ee

  To obtain the above representations we have used the relation $\rho(-q_0,\vq) = -\rho(q_0,\vq)$ (see eqn. (\ref{oddros})), as a consequence of which it follows that $\widetilde{\sigma}_R(\omega;\vq) = \widetilde{\sigma}_R(-\omega;\vq)~;~\widetilde{\sigma}_I(\omega;\vq) = - \widetilde{\sigma}_I(-\omega;\vq)$.

  The long time behavior of the inverse Laplace transform (\ref{Goftfin}) is dominated by the complex poles  of $\widetilde{\mathcal{G}}(\omega;\vq)$ closer to the real axis.  The Green's function (\ref{Gfnu})   features   complex poles corresponding to the solution of the equation
  \be \omega^2_p = E^2_\phi(\vq) + \widetilde{\sigma}(\omega_p;\vq)\,, \label{poleG}\ee which can be solved in a perturbative expansion in $g^2$. To leading order  we replace $\omega_p = \pm E_{\phi}(\vq)$ in the argument of $\widetilde{\sigma}(\omega;\vq)$. Using the property (\ref{oddros})   we find
  \be \omega_p = \pm E_{R\,\phi}(\vq)+ i \frac{\Gamma(\vq)}{2}\,, \label{polval}\ee where
  \be E_{R\,\phi}(\vq) = E_{\phi}(\vq) + \delta E_{\phi}(\vq) \,,\label{repole} \ee with
  \be \delta E_{\phi}(\vq) =   \frac{\widetilde{\sigma}_R(E_\phi(\vq))}{2E_\phi(\vq)} =  \frac{g^2}{2\,E_{\phi}(\vq)}\, \int^\infty_{-\infty} \mathcal{P}\Bigg[ \frac{\rho(q_0,q)}{(E_{\phi}(\vq)-q_0)} \Bigg]\,\frac{dq_0}{2\pi}   \,,  \label{delE}\ee and
  \be \Gamma(\vq) =  g^2\, \frac{\rho(E_\phi(\vq))}{2E_\phi(\vq)} \,,  \label{impole}\ee these are precisely the long time limits of the energy renormalization and decay rates (\ref{renfreq},\ref{gamafgr}) respectively.

  Writing in the denominator of the integrand in (\ref{Goftfin}) $\widetilde{\sigma} ( \omega;\vq) = \widetilde{\sigma} (\omega_p;\vq) + (\widetilde{\sigma}( \omega;\vq)-\widetilde{\sigma}( \omega_p;\vq))$ and expanding near the (complex) poles,  we find that near each pole, $\widetilde{\mathcal{G}}(\omega;\vq)$ can be written in
  a Breit-Wigner form as
  \be \widetilde{\mathcal{G}}(\omega;\vq) =  \frac{Z(\vq)}{2\omega_P\big(\omega \mp E_{R\phi}(\vq) -i\frac{\Gamma(\vq)}{2}\big)} \,\,, \label{BW} \ee with the quasiparticle weight or ``residue'' (wave function renormalization) given to leading order in $g^2$ by
   \be Z^{-1}(\vq) = 1- \Big[\frac{1}{2\,\omega}\,\frac{d \widetilde{\sigma}_R(\omega;\vq)}{d\omega}\Big]_{\omega = E_{\phi}(\vq)}  = 1 + \frac{g^2}{ 2E_{\phi}(\vq)}\,  \int_{-\infty}^{\infty} \,\mathcal{P}  \frac{\rho(q_0,q)}{( E_{\phi}(\vq)-q_0)^2} \, \frac{dq_0}{2\pi}   \,, \label{zwf} \ee therefore, to leading order in $g^2$ we find that $Z(\vq)$ is precisely the wave function renormalization (\ref{wfun}) obtained in the long time limit from the quantum master equation.
    To leading order in $g^2$ we find
   \be \mathcal{G}(\vq;t) = Z(\vq)\,e^{-\frac{\Gamma(\vq)}{2}t}\,\frac{\sin(E_{R\phi}(\vq)\, t)}{E_{R\phi}(\vq) }  \,. \label{goftbw} \ee In this derivation we have assumed consistently with perturbation theory,  that $\delta E_{\phi}(\vq)/E_{\phi}(\vq) \propto g^2 \ll 1$ and $\Gamma(\vq)/E_{\phi}(\vq) \propto g^2 \ll 1$ and neglected terms of higher order. The latter condition corresponds to a narrow width and long-lived quasiparticle.

   Writing the initial values in terms of complex amplitudes, we find in the long time limit

   \be \Phi(\vq;t) = Z(\vq)\,\mathcal{A}(\vq)  \, e^{-iE_{R\phi}(\vq)\, t}\, e^{-\frac{\Gamma(\vq)}{2}t} + c.c  \label{propasol1} \,,\ee where $\mathcal{A}(\vq)$ is an initial amplitude determined by $\Phi(\vq;0);\dot{\Phi}(\vq;0)$. This solution  coincides with the long time limit of the condensate from the quantum master equation (\ref{expvalqme}).

This analysis establishes a   relationship between the quantum master equation and the     many body approach to quasiparticles based on the  single particle Green's function including many body self-energy corrections. It also highlights the validity of the factorization (\ref{fact}) since to leading order in $g^2$ the self-energy is obtained from the photon bath in thermal equilibrium. However the result (\ref{propasol1}) does not reflect the time dependence of the functions $\Delta(\vq,t);\Gamma(\vq,t)$ in the Lindblad quantum master equation (\ref{Linfin}) and only captures their long time limits.   The reason for this discrepancy  is that in fact the single particle Green's function (\ref{Gfnu}) features a more complicated structure in the complex $\nu$ plane: branch cut discontinuities across the positive and negative real axis determined by the regions of support of the spectral density $\rho(q_0,\vq)$. The Breit-Wigner Lorentzian form (\ref{BW}) is only approximate and valid only in a (small) region near the complex poles which in perturbation theory are close to the real axis. While this approximation reliably captures the long time dynamics, the branch cut discontinuities are responsible for the transient dynamics.

We now present a  multi-time scale analysis of the non-local integro-differential equation (\ref{FTeom}) for the condensate that does reveal the transient dynamics. It begins with the recognition that the solutions (\ref{expvalqme},\ref{propasol1})) suggest that the spatial Fourier transform $\Phi(\vq;t)$ can be written as
\be \Phi(\vq;t) = \mathcal{A}(\vq,t)\,e^{-iE_{\phi}(\vq)\,t} + c.c \,,\label{fiamp}\ee with the amplitudes $\mathcal{A}(\vq,t)$ being \emph{slowly} varying functions of time, namely
\be \dot{\mathcal{A}}(\vq,t) \propto g^2 ~~;~~ \ddot{\mathcal{A}}(\vq,t) \ll E_{\phi}(\vq) \dot{\mathcal{A}}(\vq,t) \,.\label{slowvars}\ee Inserting the \emph{ansatze} (\ref{fiamp}) into the equation of motion (\ref{FTeom}) yields
\bea &&  e^{-iE_{\phi}(\vq)\,t}\,\Big[\ddot{\mathcal{A}}(\vq,t)-2iE_{\phi}(\vq)\,\dot{\mathcal{A}}(\vq,t)+ \int^t_0 \sigma(\vq;t-t') \, e^{iE_\phi(\vq)\,(t-t')}\,{\mathcal{A}}(\vq,t')\,dt' \Big] +     \nonumber \\
&& e^{iE_{\phi}(\vq)\,t}\,\Big[\ddot{\mathcal{A}}^*(\vq,t)+2iE_{\phi}(\vq)\,\dot{\mathcal{A}}^*(\vq,t)+ \int^t_0 \sigma(\vq;t-t') \, e^{-iE_\phi(\vq)\,(t-t')}\,{\mathcal{A}^*}(\vq,t')\,dt' \Big]=0 \,, \label{eomA}\eea and request that each bracket vanishes independently. This approximation implicitly assumes that the terms in each bracket \emph{do not} feature fast phases $e^{\pm 2i E_{\phi}(\vq)\,t}$ (note the similarity to the rotating wave approximation), we will check self-consistently the validity of this assumption in the weak coupling regime $g^2 \ll 1$.

We will study in detail the first term, since the second can be obtained from the first via the straightforward replacement $E_{\phi}(\vq) \rightarrow -E_{\phi}(\vq); \mathcal{A} \rightarrow \mathcal{A}^*$. Neglecting (self-consistently, see below) the second derivative as per the hierarchy (\ref{slowvars}), yields
\be \dot{\mathcal{A}}(\vq,t) = \frac{1}{2iE_{\phi}(\vq)}\, \int^t_0 \sigma(\vq;t-t') \, e^{iE_\phi(\vq)\,(t-t')}\,{\mathcal{A}}(\vq,t')\,dt' \,,\label{dotA}\ee since $\sigma(\vq;t-t') \propto g^2$ it follows that $\dot{\mathcal{A}} \propto g^2$ in accord with the multitime hierarchy (\ref{slowvars}). Although   equation (\ref{dotA}) can, again, be solved via Laplace transform, we invoke a separation of time scales warranted by a weak coupling and rewrite the non-local time integral as
\be \int^t_0 \sigma(\vq;t-t') \, e^{iE_\phi(\vq)\,(t-t')}\,{\mathcal{A}}(\vq,t')\,dt' \equiv \int^t_0 \frac{d}{dt'}\Big[\int^{t'}_0 \sigma(\vq;t-t'') \, e^{iE_\phi(\vq)\,(t-t'')}\,dt''\Big] \,{\mathcal{A}}(\vq,t')\,dt'\,,\label{trick}\ee integrating by parts on the right hand side and neglecting the term $\simeq \dot{\mathcal{A}}\,\sigma \propto g^4$ to leading order in $g^2$  equation (\ref{dotA}) becomes local, namely
\be \dot{\mathcal{A}}(\vq,t) = \Big[\frac{1}{2iE_{\phi}(\vq)}\, \int^t_0 \sigma(\vq;t-t') \, e^{iE_\phi(\vq)\,(t-t')}\, \,dt'\Big] \,{\mathcal{A}}(\vq,t)\,. \label{dotAloc} \ee

Notice the similarity between this treatment and the analysis based on equations (\ref{incha},\ref{incha2})  leading to the Markov approximation (\ref{Linbladmarkov}), indeed both lead to a time local equation and hinge on a separation of time scales warranted by   weak coupling.

With $\sigma(\vq;t-t')$ given by equation (\ref{SigmaFT}) the time integral yields
\be  \frac{1}{2iE_{\phi}(\vq)}\, \int^t_0 \sigma(\vq;t-t') \, e^{iE_\phi(\vq)\,(t-t')}\, \,dt' = -i \Delta(\vq,t)-\frac{\Gamma(\vq,t)}{2}\,,\label{intsig}\ee where $\Delta(\vq,t),\Gamma(\vq,t)$ are given by equations (\ref{Roftim},\ref{gamadif}) respectively,
and we obtain the final result
\be  {\mathcal{A}}(\vq,t)= e^{-i \int^t_0 \Delta(\vq,t')\,dt'}\,e^{-\frac{1}{2}\int^t_0 \Gamma(\vq,t')\,dt'}  \, {\mathcal{A}}(\vq,0) \,.\label{bigAt}\ee   Using that $\rho(-q_0,\vq) = -\rho(q_0,\vq)$ (see equation (\ref{oddros})) it follows that $\mathcal{A}^*(\vq,t)$ is precisely the complex conjugate of (\ref{bigAt}), finally yielding  to leading order in $g^2$
\be \Phi(\vq;t) = \mathcal{A}(\vq,0)\,e^{-iE_{\phi}(\vq)\,t} \,e^{-i \int^t_0 \Delta(\vq,t')\,dt'}\,e^{-\frac{1}{2}\int^t_0 \Gamma(\vq,t')\,dt'} + c.c \,,\label{finFit} \ee which confirms the result (\ref{expvalI}) with (\ref{aoftsol})  from the quantum master equation including transient dynamics  and agrees with (\ref{expvalqme}) in the long time limit.

Now, we can self-consistently assess the assumption on the hierarchy (\ref{slowvars}),
\be \ddot{\mathcal{A}} = \Big[ -i \dot{\Delta}(\vq,t)-\frac{1}{2}\,\dot{\Gamma}(\vq,t)   \Big] \, {\mathcal{A}} + \mathcal{O}(g^4) \,,\label{ddA}\ee  $\dot{\Delta},\dot{\Gamma}$ do not feature the resonant denominators and  average out on short time scales, in fact vanishing in the long time limit  and are both $\propto g^2$ and formally  $\ll E_{\phi}(\vq)$ under the assumption of weak coupling. This analysis provides a quantum many body benchmark of the various approximations invoked in the derivation of the (QME) and justifies their validity under the main assumption of weak coupling.

\subsection{Linear Response}\label{subsec:LR}
We now obtain the equation of motion for the expectation value $\overline{\Phi}(\vx,t)$ by implementing the theory of linear response, thereby establishing a direct relation between the quantum open system approach based on the quantum master equation (\ref{Linblad}) and   methods of quantum many body physics. The Heisenberg equation of motion for the (DAQ) field obtained from the Lagrangian density (\ref{synacfin}) is

 \be \ddot{ {\phi}}(\vy,t)-\big(\vec{v}\cdot\vec{\nabla}\big)\big(\vec{v}\cdot\vec{\nabla} {\phi}(\vy,t)\big)+m^2\, {\phi}(\vy,t) =  g\, \mathcal{C}(\vx,t)\,.\label{daqheiseom}\ee

 Our objective is to obtain the equation of motion for the condensate $\overline{\phi}(\vx,t)= \mathrm{Tr} \phi(\vx,t) \,\rho(0)  $, hence it is convenient to write
 \be \phi(\vx,t) \equiv \overline{\phi}(\vx,t) + \delta(\vx,t)\,,\label{daqsplit}\ee where $\delta(\vx,t)$ is a Heisenberg field operator such that by definition obeys the constraint
 \be \mathrm{Tr} \delta(\vx,t) \,\rho(0) =0 \,.\label{zerofluc}\ee

 The equation of motion (\ref{daqheiseom}) now becomes
  \be \ddot{ \overline{\phi}}(\vy,t)-\big(\vec{v}\cdot\vec{\nabla}\big)\big(\vec{v}\cdot\vec{\nabla} \overline{\phi}(\vy,t)\big)+m^2\, \overline{\phi}(\vy,t)+ \Big[\ddot{ \delta}(\vy,t)-\big(\vec{v}\cdot\vec{\nabla}\big)\big(\vec{v}\cdot\vec{\nabla} \delta(\vy,t)\big)+m^2\, \delta(\vy,t) \Big]  =  g\, \mathcal{C}(\vy,t)\,.\label{shifeom} \ee

  The interaction Hamiltonian (\ref{HIoft2}) now becomes
  \be H_i = - g\,\int d^3x \,  \Big[\overline{\phi}(\vx)+\delta(\vx)\Big]\,\mathcal{C}(\vx) \,,\label{shifHi}  \ee where $\overline{\phi}$ is a c-number that acts as an ``external source''--or ``pump''-- coupled to the Chern-Simons density $\mathcal{C}=\vec{E}\cdot\vec{B}$. Now passing to the interaction picture with unitary time evolution operator
  \be U(t) \simeq 1+ i \, g\, \int^t_0 \int d^3x \,dt' \,  \Big[\overline{\phi}(\vx,t')+\delta_I(\vx,t')\Big]\,\mathcal{C}_I(\vx,t') + \cdots \,,\label{Ushi}\ee yielding for the Heisenberg composite operator $\mathcal{C}(\vx,t)$ (see eqn. (\ref{intheis})) on the right hand side of the equation of motion (\ref{shifeom}),
  \be \mathcal{C}(\vy,t) = \mathcal{C}_I(\vy,t) + i\,g\, \int^t_0 \int   \Big[\mathcal{C}_I(\vy,t),\mathcal{C}_I(\vx,t')\Big]\,\Big[\overline{\phi}(\vx,t')+\delta_I(\vx,t')\Big] \,d^3 x dt' + \cdots\,.\label{newCi}\ee Inserting this expression into the right hand side of (\ref{shifeom}), and taking the expectation value of the resulting equation
  of motion in the initial density matrix, using that $\underset{\{\gamma\}}{\text{Tr}}\mathcal{C}_I(\vx,t)\,\rho_\gamma=0$, because the operator $\mathcal{C}_I$ is odd under parity and time reversal along with the constraint (\ref{zerofluc}), we find that the equation of motion for the expectation value $\overline{\phi}$ is precisely given by (\ref{reteom}) with the one-loop  self-energy (\ref{commu}).

  The main point of this alternative derivation of the equation of motion is to establish a direct link between the open quantum system approach based on the quantum master equation in Lindblad form and    methods from quantum many body theory, in particular linear response.

 \subsection{Induced Chern-Simons condensate:}

\begin{figure}[h]
  \centering
  \includegraphics[width=2.5in]{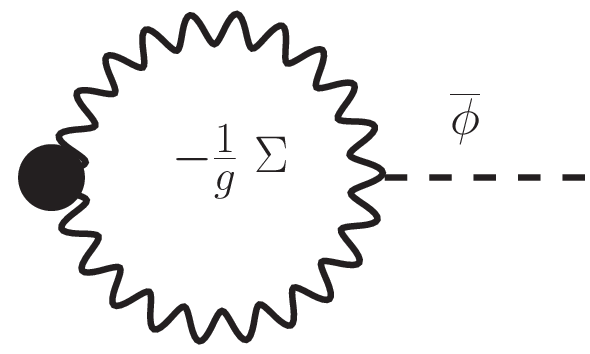}
  \caption{Expectation value of $\vec{E}\cdot \vec{B}$ (black dot) induced by the (DAC) condensate $\overline{\phi}$ (dashed line). The photon loop is $-\frac{1}{g}\Sigma$. }
  \label{fig:csexpval}
\end{figure}

 The linear response analysis above leads to an important corollary: taking the expectation value of equation (\ref{newCi}) yields to linear order in $\overline{\phi}$
 \be  \langle \mathcal{C}(\vy,t) \rangle = - \frac{1}{g}   \int^t_0 \int   \Sigma(\vy-\vx,t-t')\,\overline{\phi}(\vx,t') \,d^3 x dt' \,,\label{exCS}\ee  where $\overline{\phi}(\vx,t)$ is the solution of the equation of motion (see equation (\ref{reteom})), which to lowest order in the coupling is the solution of the free field equation of motion (\ref{expealp}). In other words, the topological Chern-Simons density acquires an expectation value induced by  a (DAQ) condensate, and the response function, namely the topological susceptibility is proportional to the (DAQ) self-energy  as  depicted by the  Feynman diagram in figure(\ref{fig:csexpval}).

This result has been previously obtained within a different context in ref.\cite{CSexp},  and  cannot be obtained \emph{directly} from the quantum master equation, as the photon degrees of freedom have been traced out and the reduced density matrix only describes the
dynamics of the (DAQ) degrees of freedom. However it is a direct corollary of the \emph{equivalence} between the dynamics of the (DAQ) as obtained from the reduced density matrix
and linear response, a bonus of the relation between the Lindblad formulation and   quantum many body physics.

The linear response equation (\ref{exCS}) can be translated to    frequency and momentum space $(\omega;\vq)$, by first taking the Laplace transform and then analytically continuing the Laplace transform variable into the complex frequency plane $s = i(\omega -i\varepsilon)$ as in the analysis of the equation of motion  (\ref{reteom}) leading up to equation (\ref{Gfnu}) in frequency and momentum space. The linear response relation (\ref{exCS}) for the space-time Fourier transforms  of the Chern-Simons density ($\widetilde{\mathcal{C}}(\omega,\vq)$) and (DAQ) condensate ($ \widetilde{\Phi}(\omega,\vq)$) becomes
\be \widetilde{\mathcal{C}}(\omega,\vq) = \chi(\omega,\vq) \,\widetilde{\Phi}(\omega,\vq) \,,\label{lrft}\ee  where
\be \chi(\omega,\vq) = -\frac{1}{g}\,  \widetilde{\sigma} (\omega;\vq)   =    -g\,\int^{\infty}_{-\infty}  \,  \frac{\rho(q_0,\vq)}{\omega-q_0-i\epsilon} \,\frac{dq_0}{2\pi}    \,,\label{suze}\ee is the susceptibility whose real and imaginary parts satisfy the Kramers-Kronig relation. In this case $\chi$ is a \emph{topological} susceptibility since the Chern-Simons density is a topological quantity.  The above relations are some of the important results of this study: the equation of motion for a (DAQ) condensate obtained from the quantum master equation before the Markov and rotating wave approximations is equivalent to that obtained from many body linear response including the retarded self-energy. Furthermore, a coherent (DAQ) condensate induces a Chern-Simons condensate, the linear (topological) susceptibility is directly related to the (DAQ) self-energy.

\section{Discussion:}\label{sec:discussion}

\begin{figure}[ht]
  \centering
  \includegraphics[width=4.5 in]{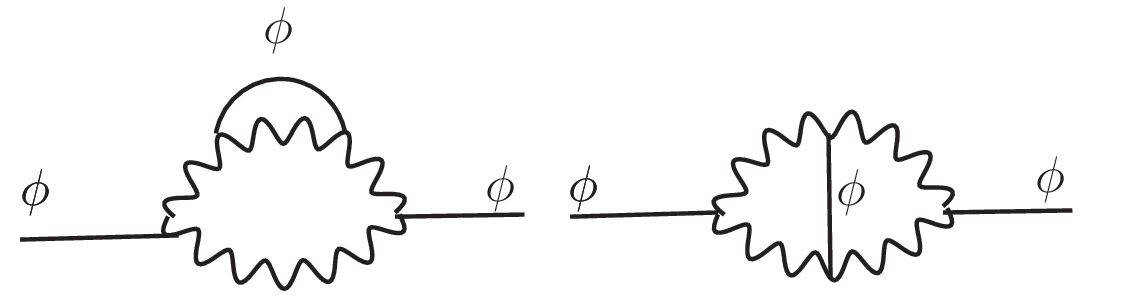}
  \caption{Correlation corrections to order $g^4$. These diagrams yield a contribution of $\mathcal{O}(g^4)$ to the equation of motion of the coherent (DAQ) condensate. }
  \label{fig:fifocorr}
\end{figure}

\textbf{System-bath Correlations:}
An important approximation invoked in obtaining the (QME) is that even during time evolution the reduced density matrix and the bath density matrix factorize, this is explicit in eqn. (\ref{fact}), an approximation that is independent of both the Markov and rotating wave approximation. As argued in ref.\cite{linme} it is expected that time evolution would induce system (DAQ)-bath correlations yielding instead the form (\ref{corre}).  It is at this stage that the correspondence between the equation of motion non-local in time for the condensate obtained from the reduced density matrix in factorized form (\ref{fact}) without further approximations (Markov and rotating wave), and the alternative derivation   based on  many body   linear response   proves illuminating. The time non-local term in the equation of motion (\ref{reteom}) is identified with the retarded one-particle irreducible self energy shown in fig.(\ref{fig:fise}) which is obtained from the \emph{free field} correlation function of the Chern-Simons density in  the free field density matrix of a photon bath in thermal equilibrium (see appendix (\ref{app:ebcoup})). Therefore the assumption of factorization up to $\mathcal{O}(g^2)$ is validated by the many body interpretation of the equation of motion. Indeed the Feynman diagram displayed in fig.(\ref{fig:fise}) is the manifestation of this factorization as there are no correlations between the photons in the bath and the (DAQ) field in the irreducible self-energy. Based on this analysis, we can now provide a Feynman diagram-based interpretation of the correlation contribution (\ref{corre}) suggested in ref.\cite{linme}. The $\mathcal{O}(g^2)$ corrections to the irreducible self-energy imply the (DAQ)-photon correlations in the self-energy shown in fig.(\ref{fig:fifocorr}), these diagrams yield contributions of $\mathcal{O}(g^4)$ to the equation of motion of the (DAQ) condensate. This is yet another bonus of establishing a relation between the  (QME) and many body theory, allowing us to understand the validity of the assumption on factorization as well as the build up of correlations in higher orders of the (DAQ)-photon coupling.

\vspace{1mm}

\textbf{Multi time correlation functions: Quantum regression. } Our analysis of the equation of motion, coherence and population dynamics only focused on single time expectation values or correlation functions. In order to obtain multi-time correlations, the (QME) must be appended with the quantum regression theorem\cite{breuer,zoeller,carmichael} which introduces the time evolution operator inserted into a chain of operators at different times. Our main objectives only required the study of single time expectation values, and while certainly multitime correlation functions are very important, they will be relegated to future study.

\vspace{1mm}

\textbf{On $\Lambda_{\max};\Lambda_p$:} The analysis in the previous sections introduced two cutoffs $\Lambda_{max};\Lambda_p$, however the effective theory defined by   the effective action (\ref{synac}) does not feature any intrinsic scale (beyond the (DAQ) effective mass) that would determine these cutoffs. The main observation is that the effective theory has been derived by ``integrating out'' (or tracing over) electronic degrees of freedom that couple to the electromagnetic fields\cite{xiao,rundong,jing,nomura,gos,mottola}. The \emph{low energy} effective action  (\ref{synac}) with a local   coupling to the Chern-Simons density is only valid below the energy scale at which the electronic degrees of freedom can be directly excited by the (DAQ) field.  Therefore, we interpret the higher energy cutoff $\Lambda_{max}$ as the scale below which the effective field theory of the (DAQ) field described by (\ref{synac}) is valid. The actual value of this upper frequency cutoff is material dependent,   for example in 3D topological insulators it is determined by the bulk band gap, typically in the range $\simeq 10^{-2}-10^{-1} \,\mathrm{eV}$, however this is not relevant for our discussion, since all the effects associated with this high frequency region are either averaged out (the oscillatory terms) or included in the (cutoff dependent) quasiparticle residue.

  The study above shows that the quasiparticle ``forms'' on a time scale $t_{qp}$ associated with the upper cutoff $\Lambda_{max}$ which under the assumption of validity of the effective field theory is well separated from the relaxation time scales, for example associated with the rate from Fermi's golden rule. The dynamics of the relaxation of the \emph{already formed quasiparticle} is described by the lower cutoff $\Lambda_p$ as displayed in figures (\ref{fig:Ipall},\ref{fig:iplus}) and is insensitive to the higher scale $\Lambda_{max}$.

The lower cutoff scale $\Lambda_p$ is not   related to $\Lambda_{max}$, it is a much lower frequency scale, introduced to separate the contributions from the spectral density near the resonance at $E_{\phi}(\vq)$  which lead to the early Zeno-like behavior $\propto t^2$ followed by the linear secular growth  \emph{a la}  Fermi's Golden rule, from those that only yield oscillatory and constant contributions to the decay function at long time. Ultimately this cutoff must be associated with the bandwidth of the measurement device\cite{zeno2,zeno3}, in other words the region of the bath spectral density that is probed by the measurement.
  For example   the observation  of the coherent oscillation in 2D $MnBi2Te4$ at $\simeq 44\,\mathrm{Ghz}$ reported in reference\cite{jianq} corresponds to  the energy scale $\simeq 0.2\,\mathrm{meV}$ implying spectroscopic  resolution in the   $\mathrm{meV}$ range.

In fact,  $\Lambda_p$ has been introduced solely as a sliding scale to split the total integrals into two different regions, one contains the resonance and the other does not, but as such  nothing depends on   $\Lambda_p$. In a sense this scale acts just like a renormalization scale describing the ``low energy'' dynamics of the quasiparticle after its formation, but the ``high energy'' part (the oscillatory parts that average out as well as the quasiparticle residue) depends on both the physical scale $\Lambda_{max}$ associated with the limit of validity of the effective field theory
and $\Lambda_p$.  Our interpretation of $\Lambda_p$ as the region of the spectral density probed by a particular  measurement of the  ``low energy'' (long time) dynamics of the quasiparticle is consistent with the interpretation as a renormalization scale.

\vspace{1mm}

\textbf{Counterrotating terms:}

In the derivation of the  quantum master equation (\ref{Linfin}) we  neglected terms of the form

\be a^\dagger(\vq)~a^\dagger(-\vq)~ e^{2iE_{\phi}(\vq) t'}~e^{i{E_{\phi}(\vq)(t-t')}}~~;~~ a(\vq)~a(-\vq)~ e^{-2iE_{\phi}(\vq) t'}~e^{-i{E_{\phi}(\vq)(t-t')}}\,. \nonumber \ee

  Their effect can be assessed  by including them into the (QME) after the (partial) Markov approximation. The time integral over $t'$ can be carried out following the steps leading to equation (\ref{Linfin}) yielding contributions of the form
$a_{\vq}\,a_{-\vq}\,e^{-2iE_{\phi}(\vq) t} \rho^\lessgtr(q_0,q) \hat{\rho}_{Ia}(t)$ etc. The contribution of these terms to the equations of motion for linear or bilinear forms of $aa^\dagger$ are straightforward to obtain. Although the equations of motion do not close in a simple manner as without these terms, they do not lead to  secular  growth in time because of the rapid oscillations  leading to rapid dephasing. These extra contributions result in non-resonant corrections   and yield perturbatively small subleading contributions    in weak coupling, as compared to those obtained from equation (\ref{Linfin}) which captures the secular growth in time because of the resonances and describes the leading behavior in the long time dynamics. Note that whereas there certainly are oscillations with
frequencies $\simeq E_{\phi}(\vq)$ in both $\mathcal{I}_+(t)$ (see fig. (\ref{fig:iplus})) and in $\mathcal{N}_p(t)$ (see fig. (\ref{fig:pop1})) these are part of the behavior that eventually leads to secular growth in time, which is not the case for the counterrotating terms. The multitime analysis of the equation of motion obtained without the rotating wave approximation which reproduced the results of the (QME) after such approximation suggests that at least to $\mathcal{O}(g^2)$ the counterrotating terms can be safely neglected.

\vspace{1mm}

  \textbf{Main differences with other systems:} The quantum master equation has been implemented to describe relaxation in a variety of systems. There are major differences between this work and most previous applications:

\textbf{i:)} whereas typically the rates in the Lindblad dissipator are either phenomenologically introduced, or calculated via Fermi's golden rule, in both cases this amounts to a Markov approximation. In contrast, our  treatment obtains the \emph{time dependent rates} from the spectral properties of the bath, their time dependence yields fully non-Markov dynamics, which translate into early time transient phenomena exhibiting violation of Fermi's golden rule and detailed balance and  anti-zeno enhancement of relaxation.

\textbf{ii:)} The time dependence of the rates unambiguously yield the  detailed dynamics associated with the build-up of the quasiparticle on time scales determined by the bandwidth of the bath, and reveals the hierarchy of time scales associated with renormalization and relaxation effects.

\textbf{iii:)} We established a direct bridge with many body physics via linear response which allows to interpret and validate many of the approximations in the Lindblad QME (seldom discussed in the literature). In particular, obtaining the equations of motion of the (DAQ) and its solution via both the (QME) and  many body linear response yields a direct link with the many body Green's function and self-energy of the (DAQ) field,  and unequivocally lead to   the relation between the topological susceptibility and the (DAQ) retarded self-energy.  The main advantage in the axion case is that there is a well defined effective quantum field theory that offers the alternative avenue of many body physics which complements and enhances the treatment with the (QME).

\vspace{1mm}

\textbf{Caveats:} Our study focused on the non-equilibrium dynamics of (DAQ) solely coupled to the electromagnetic field (the bath) via Chern-Simons term. The effective theory described by the action (\ref{synac}) contains no other degrees of freedom. In particular, as discussed above, the original electronic degrees of freedom where ``integrated out'' which is the origin of the Chern-Simons coupling in the effective field theory. It is possible that the (DAQ) could couple to other degrees of freedom, such as phonons or other collective excitations depending on
the particular topological material.
  Since the axion is a pseudoscalar field, there are certain symmetry constraints that an effective quantum field theory must address to include couplings of the axion to other degrees of freedom: if the coupling is linear in the axion field, its coupling to the other degrees of freedom must be to a pseudoscalar operator associated with them, if the coupling is bilinear, no such constraint is necessary.   Our study has nothing to say about the influence of these other degrees of freedom as it hinges on the effective action  (\ref{synac}). Our focus is on studying the non-equilibrium dynamics of (DAQ) from its coupling to electromagnetism via the topological Chern-Simons term, based on the effective action (\ref{synac}). The inclusion of other possible degrees of freedom and their effect on the non-equilibrium dynamics of (DAQ) is certainly worthy of further study, however, it is beyond the scope of our study and remains to be explored.

\vspace{1mm}

\textbf{More general lessons:} Although the focus of this study is to understand non-equilibrium aspects of (DAQ) coupled to photons via a Chern-Simons term as an open quantum system with the
electromagnetic field as a heat bath  in equilibrium, several aspects and results are more overarching. First, the ``partially Markovian approximation'' $\rho_{I\phi}(t') \rightarrow \rho_{I\phi}(t)$ yielding the (QME) (\ref{Linbladmarkov})  but without taking the infinite time limit in the correlators allows to explore non-Markov dynamics and transient phenomena. This aspect is more transparent after the rotating wave approximation yielding the (QME) in Lindblad form (\ref{Linfin}) with the coefficient functions $\Delta ; \Gamma^{\lessgtr}$ explicit functions of time, which in the usual Markov approximation these are replaced by their infinite time limit. Secondly, allowing their time dependence opens up the window into transient dynamics, with important consequences: the dynamics of formation of the quasiparticle, violations or corrections to Fermi's Golden rule and detailed balance, and  quantum (anti) Zeno dynamics. These are all general aspects that transcend the particular model and could be of potential observational relevance as modern techniques of ultrafast interferometry and spectroscopy allow to probe non-equilibrium dynamics on ever shorter time scales\cite{cetina,viva,navon}, yielding a wealth of information both on the particular system under consideration as well as fundamental non-equilibrium aspects. The spectral density   is a particular characteristic of the correlation functions of the Chern-Simons density in the photon bath that may be amenable of being studied with the analysis of transient dynamics with ultrafast techniques. Furthermore, although particular to the case under study, the connection between the open quantum system and (QME) approach with quantum many body physics, if available in other systems, may prove fruitful to understand the validity of approximations  and to  provide complementary insight into the dynamics.

\section{Conclusions and further questions:}\label{sec:conclusions}

Motivated by the recent observation of coherent oscillations interpreted as  emergent dynamical axion quasiparticles in two dimensional ($MnBi_2Te_4$)\cite{jianq}, in this article we study non-equilibrium aspects of (DAQ) coupled to electromagnetic fields via a Chern-Simons term in three dimensions as an open quantum system, treating the electromagnetic field as a bath in equilibrium.  We derive a quantum master equation for the reduced density matrix of the (DAQ) degrees of freedom invoking only a ``partial'' Markov approximation, allowing an explicit time dependence of the coefficient functions in the Lindblad form of the (QME). These are determined by the equilibrium correlation functions of the Chern-Simons density. This dependence allows us to explore non-Markovian (memory) processes and transient dynamics in the time evolution of a coherent (DAQ) condensate, coherences and population (occupation number) kinetics. We find important transient phenomena: the formation dynamics of the quasiparticle, early time evolution with enhanced quasiparticle decay, namely the  quantum anti-Zeno effect that is also imprinted in population kinetics, early time violations of Fermi's Golden rule and \emph{detailed balance}. These may be amenable to be probed by ultrafast interferometry and/or spectroscopy. A wide separation of time scales emerges as a particular characteristic of the spectral density of the bath, the shortest time scale, determined by the spectral bandwidth, determines the time scale of formation of the quasiparticle, while its decay is captured by an effective time dependent decay function on longer time scales. It features early quantum anti-Zeno behavior merging at long time with the usual linear secular growth in time, implying transient violations of Fermi's Golden rule, which are also  imprinted in the population kinetics as a transient violation of detailed balance.

We obtain the time evolution of the (DAQ) condensate from the time non-local (QME) without the Markov approximation, the non-locality in the equation of motion is recognized in terms of the quantum many body (DAQ) retarded self-energy, and is confirmed by implementing many body linear response, which affords a Feynman diagram interpretation of the main approximations in the (QME), such as factorization and emergence of system-bath correlations at higher orders. This analysis provides a bridge between quantum open systems and quantum  many body methods, a corollary of which is the recognition that the Chern-Simons density acquires an expectation value \emph{induced} by a coherent parity and time reversal breaking (DAQ) condensate, the topological susceptibility (linear response) is determined by the (DAQ) self-energy.  The transient dynamics informs on spectral characteristics of the Chern-Simons correlation functions, which may be amenable of being probed with ultrafast interferometry and/or spectroscopic techniques.

While the focus is on the non-equilibrium dynamics of (DAQ), we draw more general lessons on transient non-equilibrium dynamics that are missed by the full Markov approximation and that maybe applicable to other quantum open systems. There are important aspects that have not been explored in this study, such as the coupling of (DAQ) to other degrees of freedom that are not accounted for in the effective theory of (DAQ) and (axion) electrodynamics. These are certainly important questions that are beyond the scope of this study and remain to be explored.

\acknowledgements
 The author  gratefully acknowledges  support from the U.S. National Science Foundation through grant     NSF 2412374.

\appendix

\section{Lehmann representation of Chern-Simons  correlation functions.}\label{app:corre}
The dynamics and dissipative processes depend on the correlation functions $G^{\lessgtr}$   of the Chern-Simons   operators  in eqn. (\ref{Linblad},\ref{ggreat},\ref{gless}).

Because the bath is in thermal equilibrium, its  initial density matrix is $\rho_\gamma =e^{-\beta H_{0\gamma}}/Tr\, e^{-\beta H_{0\gamma}}$ with $H_{0\gamma}$ the free field Hamiltonian of the electromagnetic field, is space-time translationally  invariant as well as parity and time reversal invariant.  The interaction picture operators associated with the bath are given by $\mathcal{C}_I(\vx,t) = e^{iH_{0\gamma} t}\,e^{-i\vec{P}\cdot \vx} \mathcal{C}_I(\vec{0},0)\,e^{-iH_{0\gamma} t}\,e^{i\vec{P}\cdot \vx}$, where $\vec{P}$ is the   momentum operator. Therefore,   we can write
\bea  G^>(\vx-\vx';t-t')  & = & \mathrm{Tr}\, \rho_{\gamma}\, \mathcal{C}_I(\vx,t)\, \mathcal{C}_I(\vx',t')\,   = \int \frac{d^3k}{(2\pi)^3}\int \frac{dk_0}{2\pi}~ \rho^>(k_0,\vk) e^{-ik_0(t-t')}\,e^{i\vk\cdot(\vx-\vx^{\,'})} \label{Ggfd} \\
G^<(\vx-\vx';t-t')  & = &  \mathrm{Tr}\, \rho_\gamma \, \mathcal{C}_I(\vx',t')\, \mathcal{C}_I(\vx,t)\,  = \int \frac{d^3k}{(2\pi)^3}\int \frac{dk_0}{2\pi}~ \rho^<(k_0,\vk) e^{-ik_0(t-t')}\,e^{i\vk\cdot(\vx-\vx^{\,'})} \,. \label{Glfd} \eea These representations are obtained by   introducing a complete set of simultaneous eigenstates of $H_{0\gamma}$ and the total momentum operator $\vec{P}$,  $(H_{0\gamma},\vec{P})\ket{n} = (E_n,\vec{P}_n)\ket{n}$, from which we obtain the following Lehmann representations\cite{fetter,mahan},
\begin{eqnarray}
\rho^>(k_0,\vk) & = &  \frac{(2\pi)^4}{\mathrm{Tr}e^{-\beta H_{\gamma}}}~
\sum_{m,n}e^{-\beta E_n}
|\langle n| \, \mathcal{C}_I(\vec{0},0) \, |m \rangle|^2  \, \delta(k_0-(E_m-E_n))\,\delta(\vec{k}-(P_m-P_n)) \label{siggreat} \\
\rho^<(k_0,\vk) & = &  \frac{(2\pi)^4}{\mathrm{Tr}e^{-\beta H_{\gamma}}}~
\sum_{m,n} e^{-\beta E_n}
 |\langle m| \, \mathcal{C}_I(\vec{0},0)\,  |n \rangle|^2  \, \delta(k_0-(E_n-E_m))\,\delta(\vec{k}-(P_n-P_m))\,.
 \label{sigless}
\end{eqnarray} Upon relabelling
$m \leftrightarrow n$ in the sum in the definition (\ref{sigless}) and recalling that $\mathcal{C}_I$ is an hermitian operator,
we find the Kubo-Martin-Schwinger relation\cite{kms}

\begin{equation}
\rho^<(k_0,\vk)  = \rho^>(-k_0,\vk) = e^{-\beta k_0}
\rho^>(k_0,\vk)\,. \label{KMS}
\end{equation}

  The spectral density is defined as
\be \rho(k_0,\vk) = \rho^>(k_0,\vk)-\rho^<(k_0,\vk) = \rho^>(k_0,\vk)\big[ 1-e^{-\beta k_0}\big] \label{specOs}\ee
therefore
\be  \rho^>(k_0,\vk) = \rho(k_0,\vk)~\big[1+n(k_0)\big]~~;~~\rho^<(k_0,\vk) = \rho(k_0,\vk)~ n(k_0) \,, \label{relas}\ee  where
\be n(k_0) = \frac{1}{e^{\beta k_0}-1} \,. \label{bose}\ee

Furthermore, from the first equality in (\ref{KMS}) it follows that
\bea \rho(-k_0,\vk) & = &  - \rho(k_0,\vk) \,,  \label{oddros}\\ \rho(k_0,\vk) & > & 0 ~~\mathrm{for} ~~ k_0 > 0 \,. \label{positive}
\eea

   The Lehmann representations (\ref{siggreat},\ref{sigless}) are non-perturbative and unambiguously yield the relations (\ref{KMS}-\ref{positive}) which are general, non-perturbative statements relying on thermal equilibrium and space-time translational invariance and do not depend on these couplings.

\section{Spectral density for $\vec{E}\cdot \vec{B}$ correlation functions.}\label{app:ebcoup}
The quantized  gauge field in interaction picture are given by equation (\ref{vecpot2})
  where  $\hat{\epsilon}_{\vk,\lambda}$ are the transverse polarization vectors chosen to be real and defined so that $\hat{\epsilon}_{\vk,1,2};\hat{\vk}$ form a right handed triad, namely
   \be \vec{\epsilon}_{1}(\vk)\times\vec{\epsilon}_{2}(\vk)= \hat{\vk}~~;~~\vec{\epsilon}_{2}(\vk)\times\hat{\vk}= \vec{\epsilon}_{1}(\vk)~~;~~\vec{\epsilon}_{1}(\vk)\times\hat{\vk}=-\vec{\epsilon}_{2}(\vk)~~;~~ \vec{\epsilon}_{1}(-\vk)= - \vec{\epsilon}_{1}(\vk)~~;~~\vec{\epsilon}_{2}(-\vk)= \vec{\epsilon}_{2}(\vk)\,. \label{polas}\ee

   From  eqns (\ref{Ggfd},\ref{Glfd}) we need the correlation functions
\bea G^>(x-y) & = & \langle \vec{E}(x)\cdot \vec{B}(x)\vec{E}(y)\cdot \vec{B}(y)\rangle \,,\label{Ggeb} \\
G^<(x-y) & = & \langle \vec{E}(y)\cdot \vec{B}(y)\vec{E}(x)\cdot \vec{B}(x)\rangle = G^>(y-x) \,, \label{Gleb} \eea where $\vec{E}(\vx,t),\vec{B}(\vx,t)$ are given by equation (\ref{ebfields}),   and  $\langle (\cdots ) \rangle$  are defined as averages in the thermal density matrix of free field photons.

In the thermal ensemble the expectation value $\langle \vec{E}(x)\cdot \vec{B}(x) \rangle =0$ by parity  and time reversal invariance.      Using Wick's theorem, the $(\vec{E}\cdot\vec{B})$ correlation function becomes
\be \langle \vec{E}(x)\cdot \vec{B}(x)\vec{E}(y)\cdot \vec{B}(y)\rangle = \sum_{i,j}\Big\{ \langle E^i(x)\, E^j(y) \rangle \langle B^i(x)\, B^j(y) \rangle + \langle E^i(x)\, B^j(y) \rangle \langle B^i(x)\, E^j(y) \rangle  \Big\}\,.  \label{correEB}\ee A straightforward calculation yields
\be  \langle E^i(x)\, E^j(y) \rangle  = \langle B^i(x)\, B^j(y) \rangle = \frac{1}{2V}\sum_{\vk} k\,\Big(\delta^{ij}-\hat{\vk}^i \hat{\vk}^j\Big)\,\Big[(1+n(k))\,e^{-ik\cdot(x-y)} + n(k) \, e^{ ik\cdot(x-y)}\Big] \,, \label{eecor}\ee similarly
\be \langle E^i(x)\, B^j(y) \rangle  = - \langle B^i(x)\, E^j(y) \rangle = -\frac{1}{2V} \sum_{\vk} k\, \Big(\hat{\epsilon}^i_{\vk,1}\,\hat{\epsilon}^j_{\vk,2}-\hat{\epsilon}^i_{\vk,2}\,\hat{\epsilon}^j_{\vk,1} \Big)\, \Big[(1+n(k))\,e^{-ik\cdot(x-y)} + n(k) \, e^{ ik\cdot(x-y)}\Big] \,,\label{ebcorr} \ee where $n(k) = 1/(e^{\beta k} -1)$. Combining the two terms in (\ref{correEB}) we find
\bea G^>(x-y) &  = &  \frac{1}{4V^2} \sum_{\vk}\sum_{\vp} k\,p\,(1-\hat{\vk}\cdot \hat{\vp})^2 \Bigg\{ \Big[(1+n(k))\,e^{-ik\cdot(x-y)} + n(k) \, e^{ ik\cdot(x-y)}\Big]\nonumber \\ & \times & \Big[(1+n(p))\,e^{-ip\cdot(x-y)} + n(p) \, e^{ ip\cdot(x-y)}\Big]\Bigg\}\,.  \label{Ggfi}\eea
Expanding the product, we perform the following change of variables in the various terms:

 \vspace{1mm}

 \textbf{1)} in the term $n(k)n(p)$: $\vk \rightarrow -\vk,\vp \rightarrow -\vp$;

  \vspace{1mm}

  \textbf{2)} in the term with $(1+n(k))n(p)$: $\vp \rightarrow -\vp$;

  \vspace{1mm}

  \textbf{3)} in the term with $n(k)(1+n(p))$: $\vk \rightarrow -\vk$.

    Taking the infinite volume limit  with $(1/V) \, \sum_{\vq} \rightarrow \int d^3q/(2\pi)^3 $ we obtain
\be G^>(x-y) = \int \frac{dq_0}{2\pi}\int \frac{d^3q}{(2\pi)^3}\,\rho^>(q_0,q)\, e^{-iq_0(t-t')}\,e^{i\vec{q}\cdot(\vx-\vy)}\,,\label{ggro} \ee where
\bea \rho^>(q_0,q) & = &  \frac{\pi}{2}\int \frac{d^3k}{(2\pi)^3}k |\vq-\vk|\Bigg\{\Big(1-\frac{\vk}{k}\cdot\frac{\vq-\vk}{|\vq-\vk|} \Big)^2\,\Big[(1+n(k))(1+n(|\vq-\vk|))  \delta(q_0-k-|\vq-\vk|)\nonumber \\ & + &  n(k)n(|\vq-\vk|)\,\delta(q_0+k+|\vq-\vk|)\Big]\nonumber \\
& + & \Big(1+\frac{\vk}{k}\cdot\frac{\vq-\vk}{|\vq-\vk|} \Big)^2\,\Big[(1+n(k))n(|\vq-\vk|)\delta(q_0-k+|\vq-\vk|) \nonumber \\ & +  & n(k)(1+n(|\vq-\vk|)) \delta(q_0+k-|\vq-\vk|)\Big]\Bigg\}\,.\label{roge} \eea Writing
\be G^<(x-y) = \int \frac{dq_0}{2\pi}\int \frac{d^3q}{(2\pi)^3}\,\rho^<(q_0,q)\, e^{-iq_0(t-t')}\,e^{i\vec{q}\cdot(\vx-\vy)}\,,\label{glro} \ee and using the relation (\ref{Gleb}) we find that $\rho^<(q_0,\vq) = \rho^>(-q_0,-\vq)$, however the sign change in $\vq$ can be compensated by $\vk \rightarrow -\vk$ inside the k-integral with the final result
\be \rho^<(q_0,\vq) = \rho^>(-q_0,\vq) \,, \label{iden}\ee  furthermore, using the identity $(1+n(w)) = e^{\beta w} n(w)$ and using the various delta functions in the definition of $\rho^>$ we find
\be \rho^<(q_0,\vq) = e^{-\beta q_0}\,\rho^>(q_0,\vq)\,, \label{inden2}\ee which is the Kubo-Martin-Schwinger relation\cite{kms}, thereby confirming the general results  (\ref{KMS}). The spectral density is given by (see eqn. (\ref{specOs})) $\rho(q_0,q) = \rho^>(q_0,q) - \rho^<(q_0,q)$ with
\bea && \rho(q_0,q)    =     \frac{\pi}{2}\int \frac{d^3k}{(2\pi)^3} \frac{1}{k w} \Bigg\{\big(k w + k^2 -\vk\cdot\vq \big)^2\,[1+n(k)+n(w)]\big(\delta(q_0-k-w)-\delta(q_0+k+w) \big) \nonumber \\ & + & \big(k w - k^2 +\vk\cdot\vq \big)^2\,(n(w)-n(k))\big(\delta(q_0-k+w)-\delta(q_0+k-w) \big)    \Bigg\}~~;~~ w = |\vq-\vk|\,.  \label{rhoeb} \eea
The spectral density is calculated by implementing the following steps:
\be \int \frac{d^3k}{8\pi^3} = \int^\infty_0 k^2 \frac{dk}{4\pi^2} d(cos(\theta)) ~~;~~ w = |\vq-\vk| = \sqrt{q^2+k^2-2kq\cos(\theta)}~~;~~ \frac{d(\cos(\theta))}{w}  = -\frac{d\,w}{kq} \,.\label{steps}\ee Carrying out the integrations, which are facilitated by the delta function constraints we find

\begin{equation}
    \rho(q_0,\vec{q})
    = \frac{(Q^2)^2}{32\pi}\,\Bigg\{\Bigg(1 + \frac{2}{\beta q}\,\ln\Bigg[\frac{1-e^{-\beta \omega^I_+}}{1-e^{-\beta \omega^I_-}} \Bigg]\Bigg)\,\Theta(Q^2)  + \frac{2}{\beta q}\, \ln\Bigg[\frac{1-e^{-\beta \omega^{II}_+}}{1-e^{-\beta \omega^{II}_-}} \Bigg]\,\Theta(-Q^2) \Bigg\}\, \mathrm{sign}(q_0)\,, \label{rhofi}
\end{equation} where
\be Q^2= q^2_0 - q^2 ~~;~~ \omega_\pm^{I} = \frac{|q_0| \pm q}{2}~~;~~ {\omega}_\pm^{II} = \frac{q \pm |q_0|}{2}\,.\label{Q2omegas}\ee

Two limits in which the spectral density simplifies are of particular importance:

\vspace{1mm}

\textbf{a)  long wavelength limit $q \rightarrow 0$:} in this limit the spectral density simplifies to
\be   \rho(q_0,\vec{0})
    = \frac{q^4_0}{32\pi}\,\Big[1+2 \,n\Big(\frac{\beta |q_0|}{2} \Big) \Big]\, \mathrm{sign}(q_0) \,, \label{rhoqzero}\ee
    \vspace{1mm}

    \textbf{b) zero temperature limit:} in this case the spectral density is given by
    \be \rho(q_0,\vec{q})
    = \frac{(q^2_0-q^2)^2}{32\pi}\,\Theta(q^2_0-q^2)\,\mathrm{sign}(q_0)\,. \label{zeroTrho} \ee

\end{document}